\documentstyle[psfig,onecolumn]{mn}
\newif\ifAMStwofonts

\newcommand{\etal}{et al. }

\newcommand{\adhoc}{{ad hoc} }

\newcommand{\mytorus}{{MYTORUS} }
\newcommand{\mytorusp}{{MYTORUS}}

\newcommand{\rxte}{{\it RXTE} }

\newcommand{\asca}{{\it ASCA} }

\newcommand{\xmm}{{\it XMM-Newton} }
\newcommand{\xmmp}{{\it XMM-Newton}}
\newcommand{\chandra}{{\it Chandra} }

\newcommand{\suzaku}{{\it Suzaku} }
\newcommand{\suzakup}{{\it Suzaku}}

\newcommand{\bsaxp}{{\it BeppoSAX}}
\newcommand{\swift}{{\it Swift} }
\newcommand{\swiftp}{{\it Swift}}
\newcommand{\nustar}{{\it NuSTAR} }
\newcommand{\nustarp}{{\it NuSTAR}}

\newcommand{\fekalfa}{{Fe~K$\alpha$} }
\newcommand{\fekalfap}{{Fe~K$\alpha$}}

\newcommand{\fekbeta}{{Fe~K$\beta$} }

\newcommand{\fexxv}{Fe~{\sc xxv} }

\newcommand{\fexxvres}{Fe~{\sc xxv}(r) }

\newcommand{\fexxvi}{Fe~{\sc xxvi} }

\newcommand{\feklya}{{Fe~{\sc xxvi }~Ly$\alpha$} }

\newcommand{\feklyap}{{Fe~{\sc xxvi}~Ly$\alpha$}}

\newcommand{\thetaobs}{{$\theta_{\rm obs}$} }
\newcommand{\thetaobsp}{{$\theta_{\rm obs}$}}

\newcommand{\cxispin}{$C_{\rm PIN:XIS}$ }

\newcommand{\tablectratesp}{{Table~1}}
\newcommand{\tablemytfits}{{Table~2} }
\newcommand{\tablemytfitsp}{{Table~2}}

\newcommand{\srcname}{{Fairall~9} }
\newcommand{\srcnamep}{{Fairall~9}}

\ifoldfss
  \ifCUPmtlplainloaded \else
    \NewTextAlphabet{textbfit} {cmbxti10} {}
    \NewTextAlphabet{textbfss} {cmssbx10} {}
    \NewMathAlphabet{mathbfit} {cmbxti10} {} 
    \NewMathAlphabet{mathbfss} {cmssbx10} {} 
  \fi
  \ifAMStwofonts
    \ifCUPmtlplainloaded \else
      \NewSymbolFont{upmath} {eurm10}
      \NewSymbolFont{AMSa} {msam10}
      \NewMathSymbol{\upi}     {0}{upmath}{19}
      \NewMathSymbol{\umu}     {0}{upmath}{16}
      \NewMathSymbol{\upartial}{0}{upmath}{40}
      \NewMathSymbol{\leqslant}{3}{AMSa}{36}
      \NewMathSymbol{\geqslant}{3}{AMSa}{3E}

    \fi
  \fi
\fi 

\ifCUPmtlplainloaded \else
  \ifAMStwofonts \else 
    \def\upi{\pi}
    \def\umu{\mu}
    \def\upartial{\partial}
  \fi
\fi

\title{No Signatures of Black-Hole Spin in the X-ray Spectrum of the Seyfert~1 Galaxy Fairall~9}
\author[T. Yaqoob \etal]{T. Yaqoob$^{1,2}$, T. J. Turner$^{2}$, M. M. Tatum$^{2}$, M. Trevor$^{2}$, A. Scholtes$^{2}$ \\
	$^1$Department of Physics and Astronomy, Johns Hopkins University, 3400 N. Charles St., Baltimore, MD 21218. \\
	$^2$Department of Physics, University of Maryland Baltimore County, Baltimore, MD 21250.\\
}
\date{
      Received }

\pagerange{\pageref{firstpage}--\pageref{lastpage}}
\pubyear{2016}

\begin{document}

\maketitle

\label{firstpage}

\begin{abstract} 

Fairall~9 is one of several type 1 active galactic nuclei for which it has been claimed that the angular momentum (or spin) of the supermassive black hole can be robustly measured, using the \fekalfa emission line and Compton-reflection continuum in the X-ray spectrum. The method rests upon the interpretation of the \fekalfa line profile and associated Compton-reflection continuum in terms of relativistic broadening in the strong gravity regime in the innermost regions of an accretion disc, within a few gravitational radii of the black hole. Here, we re-examine a \suzaku X-ray spectrum of Fairall~9 and show that a face-on toroidal X-ray reprocessor model involving only nonrelativistic and mundane physics provides an excellent fit to the data. The \fekalfa line emission and Compton reflection continuum are calculated self-consistently, the iron abundance is solar, and an equatorial column density of $\sim 10^{24} \ \rm cm^{-2}$ is inferred. In this scenario, neither the \fekalfa line, nor the Compton-reflection continuum provide any information on the black-hole spin. Whereas previous analyses have assumed an infinite column density for the distant-matter reprocessor, the shape of the reflection spectrum from matter with a finite column density eliminates the need for a relativistically broadened \fekalfa line. We find a 90 per cent confidence range in the \fekalfa line FWHM of $1895$--$6205 \ \rm km \ s^{-1}$, corresponding to a distance of $\sim 3100$ to $33,380$ gravitational radii from the black hole, or $0.015$--$0.49$~pc for a black-hole mass of $\sim 1-3 \times 10^{8} \ M_{\odot}$.

\end{abstract}


\section{Introduction}
\label{intro}

Fairall~9 is a bright, nearby ($z=0.047016$) Seyfert~1 galaxy that has been the subject of several studies
that attempt, using X-ray spectroscopy, to obtain robust measurements of the angular momentum, or spin, 
of the putative supermassive central black hole (Schmoll \etal 2009; 
Emmanoulopoulos \etal 2011; Patrick \etal 2011a, 2011b, 2013; Lohfink \etal 2012a; Walton \etal 2013; Lohfink \etal 2016). 
According to the ``no hair'' theorem, black holes
have only three measurable physical attributes, one of these being spin (mass and charge being the other
two). In addition to its importance for fundamental physics, black-hole spin affects the energetics and
evolution of the environment in which the black hole resides. Thus, the spin properties of black holes
in a population of sources, such as active galactic nuclei (AGN), could provide clues pertaining to the
formation and growth of supermassive black holes (e.g. see Volonteri \& Begelman 2010).

The reason why Fairall~9 has been of particular interest for black-hole spin measurements via X-ray spectroscopy
is that it is a member of a subset of AGN that have been established to exhibit few or no signatures
of line-of-sight absorption that could complicate modeling of the X-ray spectral features that are thought
to carry signatures of black-hole spin (Gondoin \etal 2001; Emmanoulopoulos \etal 2011). 
Otherwise known as ``bare Seyfert galaxies'', AGN such as Fairall~9
provide a ``clean'' direct view of the accreting black-hole system (e.g. Patrick \etal 2011a; 
Tatum \etal 2012; Walton \etal 2013). It is the X-ray spectrum from the
innermost regions of the accretion disc that is thought to convey the signatures of relativistic effects
in the vicinity of the black hole, including its spin. 
The basic method for constraining the black-hole spin involves fitting a model of the X-ray reflection
spectrum from the accretion disc, and the associated \fekalfa emission line, with the broadening,
or ``blurring'' effects of Doppler and gravitational energy shifts being the key drivers
(e.g., see Reynolds 2014; Middleton 2015, and references therein). The disc-reflection model invariably
used in the literature for AGN in general is REFLIONX (Ross, Fabian \& Young 1999; Ross \& Fabian 2005),
combined with one of several
alternatives for the ``blurring'' kernel 
(e.g. Laor 1991; Dov\v{c}iak, Karas \& Yaqoob 2004; Brenneman \& Reynolds 2006; Dauser \etal 2010). 
Other model components often need to be included to
account for features that may or may not be directly related to the accreting black-hole system
(for example, additional or alternative soft X-ray emission, and/or narrow emission lines from distant matter
at tens of thousands of gravitational radii from the black hole).
Clearly, the fewer the number of model components that are needed, the less ambiguity and/or degeneracy there will
be in constraining the black-hole spin. The reason why \suzaku spectra for Fairall~9 have received
the attention of several studies is that the good energy resolution of the CCD detectors for 
measuring the crucial \fekalfa line profile, combined with the simultaneous 
sensitive coverage above 10~keV, provides a significant advance over previous capabilities for
constraining the X-ray reflection spectrum and fluorescent line emission.
Early observations of Fairall~9 with \asca (Reynolds 1997) and \xmm 
(Gondoin \etal 2001; Emmanoulopoulos \etal 2011) lacked the high-energy coverage. Nevertheless,
a broad \fekalfa line (the principal feature required for constraining black-hole spin)
was reported for the \asca observation (Reynolds 1997) and for a 2009 \xmm observation
(Emmanoulopoulos \etal 2011), but for an earlier \xmm observation in 2000, Gondoin \etal (2001) reported
no detection of a broad \fekalfa line. More recently, Fairall~9 was targeted by a \swift monitoring
campaign and three new \xmm observations, one of them simultaneous with \nustar (Lohfink \etal 2016). 
Although it was stated by the authors that they ``clearly detect blurred ionised reflection,'' 
detection of a relativistically broadened line alone (as opposed to the joint detection of such a line
and the reflection continuum as a single spectral component), was not claimed.

The result of all the X-ray spectroscopy studies of Fairall~9 that attempt to measure the
black-hole spin is a wide range in its inferred value, ranging from zero to 0.998, the theoretical
maximum for accretion (Thorne 1974). Analyses of different data sets, as well as the same data set with
different models, yield different black-hole spin measurements, some of which are inconsistent
within the given errors. In some cases, the inferred inclination angle of the accretion disc
and the iron abundance are also inconsistent amongst different data sets and models applied. 
Lohfink \etal (2012a) tried to resolve these serious conflicts by adding an additional soft X-ray
continuum to their model and forcing the black-spin, disc inclination angle, and iron abundance
to be invariant amongst different \xmm and \suzaku data sets. However, this procedure involved
adding nine free parameters in total to an already complex model, so it is not surprising that
a consistent value of black-hole spin was able to fit the data. Moreover, Lohfink \etal (2012a)
explained that the two different models of the additional soft X-ray continuum both left 
unresolved problems. In one case the additional soft X-ray continuum was a thermal 
Comptonisation component that implied a colossal iron overabundance, with a lower limit of 8.2 
relative to solar. In the other case, the additional component was an ionised reflection spectrum 
(using the REFLIONX model), but this solution implied a radial ionisation gradient in the
accretion disc that predicts atomic features in the spectrum that are not observed.
Even if a viable consistent solution for the black-spin can be found for different data sets,
the problem of the model-dependence of the black-hole spin based on fitting the same data 
with different models
still remains. Patrick \etal (2011b) studied this model dependence for a small sample of
AGN observed by \suzaku (including Fairall~9), and the severity of the model dependence
led them to conclude that ``zero spin cannot be ruled out at a particularly high confidence
level in all objects.'' The principal difference in the applied models pertained to the
exact model used for relativistic broadening of the \fekalfa line (including the option of omitting
that component altogether). Several choices are available for such a model in popular 
X-ray spectral-fitting packages (e.g. Dov\v{c}iak \etal 2004; Dauser \etal 2010; Patrick \etal 2011a; Middleton 2015). 

It is also known that partial covering models can sometimes
eliminate the need for a relativistically broadened \fekalfa emission line (e.g. Miller, Turner \& Reeves 2008; Iso \etal 2016).
In such a scenario, there is degeneracy between the spectral curvature due to
relativistic broadening of the \fekalfa line and the spectral curvature due to absorption by
matter that partially covers the X-ray source.
However, only one-dimensional partial covering models are typically
applied, which may be incomplete because only line-of-sight
absorption is considered, and Compton scattering and fluorescent line emission from any globally
distributed matter is not included. Recently, a more sophisticated partial covering model has been
applied to Mkn~335 (Gallo \etal 2015), but the spectra were so complicated that both
the blurred reflection model and the partial covering model left unmodeled residuals. 
It is therefore important to first understand the much simpler spectra of bare Seyfert galaxies such as Fairall~9.
Another type of model that 
eliminates the need for a relativistically broadened \fekalfa line whilst also accounting
for matter out of the line of sight invokes a Compton-thick accretion disc wind (Tatum \etal 2012),
and has also been successfully applied to Fairall~9.
A different study involving a disc wind model, applied to the Narrow Line Seyfert galaxy 1H~0707--495 
(Hagino \etal 2016), also led to the conclusion that the X-ray spectrum in this source can be modeled with absorption in the wind
rather than invoking relativistic effects in strong gravity to produce a broad \fekalfa line.
The implication of these studies is that if there is no relativistically broadened \fekalfa line emission
from the innermost regions of the accretion disc, then there is no signature of black-hole spin
in the X-ray spectrum. In this scenario, measurements of black-hole spin obtained from
applying models that do have relativistically broadened \fekalfa line emission are then
artifacts of incorrect modeling.

One feature that is common to all of the above-mentioned models fitted to the X-ray spectra of
Fairall~9 is the method of modeling the narrow core of the \fekalfa line, which is ubiquitous in 
both type~1 and type~2 AGN (e.g. Yaqoob \& Padmanabhan 2004; Nandra 2006; Shu, Yaqoob \& Wang 2010, 2011;
Fukazawa \etal 2011; Ricci \etal 2014). Hereafter, it will be referred to
as the distant-matter \fekalfa line, since it has a width that is typically less than a few
thousand ${\rm km \ s^{-1}}$ full width half maximum (FWHM), corresponding to an origin in matter
located at tens of thousands of gravitational radii from the black hole. Shu \etal (2011) measured
a mean FWHM of $\sim 2000 \pm 160 {\rm km \ s^{-1}}$ from a sample of AGN observed by the \chandra high energy grating (HEG).
They showed that in some AGN the location of the distant-matter \fekalfa line emitter is consistent
with the classical broad line region (BLR), yet in other AGN the location is further from the black hole,
possibly at the site of a putative obscuring torus structure that is a principal ingredient of AGN
unification schemes (e.g., Antonucci 1993; Urry \& Padovani 1995). The peak energy of the narrow \fekalfa line
is tightly distributed around 6.4~keV for both type~1 and type~2 AGN 
(e.g. Sulentic \etal 1998; Yaqoob \& Padmanabhan 2004; Nandra 2006; Shu \etal 2010, 2011),
providing a strong observational constraint on the neutrality of the matter producing the line,
and any associated Compton-scattered continuum.
Any attempt to measure the relativistically broadened \fekalfa line
in AGN X-ray spectra must properly account for the narrow distant-matter \fekalfa line because it can carry 
a luminosity that is comparable to, or even greater than that in the broad component. The models
must also account for the Compton-scattered (reflection) continuum in the matter that produces the
narrow \fekalfa line. Without exception, the previous studies of Fairall~9 described above all
use a model for the narrow \fekalfa line and the associated reflection continuum that is based
on matter with an infinite column density, a flat (disc) geometry, and an infinite size,
with the inclination angle of the disc relative to the observer fixed at an arbitrary value.
In some cases the narrow \fekalfa line emission and the associated reflection continuum are calculated
self-consistently using the model PEXMON (Nandra \etal 2007) or even a second
ionised disc-reflection spectrum (e.g. REFLIONX) with the relativistic blurring turned off (e.g. Gallo \etal 2015). 
In other cases the \fekalfa line emission 
is modeled with a simple Gaussian, the reflection continuum is modeled with PEXRAV (Magdziarz \& Zdziarski 1995),
and the two components (which are in reality not independent), are given \adhoc (floating) normalizations. In yet another
variant, the narrow \fekalfa line emission is modeled with a Gaussian component and
the associated reflection continuum is completely ignored and omitted (e.g. Patrick \etal 2013).
The assumption in the latter scenario is that 
the narrow \fekalfa line core originates in Compton-thin matter and that the reflection
continuum from that Compton-thin matter is negligible and can be ignored. However, there is no basis for the
omission of the reflection continuum from Compton-thin matter and indeed, it was shown in the
case of Mkn~3 that it cannot be neglected (Yaqoob \etal 2015).
Moreover, several recent studies have shown that the Compton-scattered continuum from matter with a finite
column density is different to that from matter with an infinite column density,
and exhibits a rich variation in spectral shape (Murphy \& Yaqoob 2009; Ikeda \etal 2009;
Yaqoob 2012; Liu \& Li 2014; Yaqoob \etal 2015; Furui \etal 2016). The so-called ``Compton-hump'' 
(the peak in the reflection spectrum) is no longer located at $\sim 30$~keV but depends on
the column density (as well as geometry and inclination angle), 
and can actually be located below the energy of the core of the
\fekalfa line for column densities $<10^{24} \ {\rm cm^{-2}}$. 
The complexity of the reflection continuum from matter with a finite column density can potentially model
features that are traditionally interpreted as relativistically broadened \fekalfa line
emission. 

In this paper, we make no assumptions about the column density
of the matter producing the narrow \fekalfa line in Fairall~9 and 
use the \mytorus model of a toroidal X-ray reprocessor (Murphy \& Yaqoob 2009) 
to actually let the data determine the global column density of
the reflecting material from spectral fitting. 
In the \mytorus model, the \fekalfa line shape and flux are calculated self-consistently
with respect to the associated reflection continuum. A Compton-thick reflection
spectrum similar to that from infinite column density disc-reflection models
can be recovered as a limiting case in the \mytorus model, which is based
on a more realistic geometry for the circumnuclear matter than the disc geometry that has 
has been used in all of the previous studies of Fairall~9 (based on models such as PEXRAV and
PEXMON). In type~2 AGN, column densities in the line of sight have been observed over
a wide range, from Compton-thin to Compton-thick, with some sources exhibiting transitions
between Compton-thin and Compton-thick states 
(e.g. Matt, Guainazzi \& Maiolino 2003; Risaliti \etal 2010, and references therein). 
It should therefore not be surprising if
this range of column densities is also appropriate for the circumnuclear matter in type~1 AGN.
We note that from extended \rxte monitoring data of Fairall~9, Lohfink \etal (2012b) 
presented evidence of clumps of matter transiting across the X-ray source that did not
completely extinguish the X-ray emission. 
The column density of the clumps was crudely estimated by Lohfink \etal (2012b) to be 
``a few $\times 10^{24} \ \rm cm^{-2}$. However, Markowitz, Krumpe, \& Nikutta (2014)
pointed out that the \rxte monitoring data are also consistent with intrinsic
spectral variability, an interpretation also given in later work by Lohfink \etal (2016), 
based on a \swiftp/\xmmp/\nustar campaign on Fairall~9 in 2014.

In this paper we apply the \mytorus model to \suzaku data for Fairall~9. 
The \suzaku data are well-suited 
for studying the X-ray reprocessor because \suzaku provides simultaneous
coverage of the critical Fe~K band ($\sim 6-8$~keV) with
good spectral resolution and throughput (due to the XIS CCD detectors), and of the hard X-ray
band above 10~keV with good sensitivity. The simultaneous broadband coverage is important
for reconciling the absorbed and reflected continua with the \fekalfa line emission.
The scope and intent of the present work was to focus on \suzakup/\chandra HEG data in the
short term and provide a framework and basis for additional studies in the long term,
e.g. with \nustarp. We will show that our model, which includes no 
relativistically broadened \fekalfa line or disc reflection, gives an excellent fit to the \suzaku data. 
The X-ray spectrum of Fairall~9, which can be explained by \fekalfa line emission and 
X-ray reflection from the distant matter alone, would then carry no information on black hole spin. 

The paper is organized as follows. In \S\ref{obsdata} we describe the
basic data, and reduction procedures.
In \S\ref{analysisstrategy} we describe the analysis strategy, including the procedures for setting
up the X-ray reprocessing models
for spectral fitting. In \S\ref{spfitting}
we give the detailed results from spectral fitting.
In \S\ref{summary} we summarize our results and conclusions.

\section{\suzaku Observation and Data Reduction}
\label{obsdata}

The present study pertains to an observation of 
Fairall~9 made by the joint Japan/US X-ray astronomy satellite, \suzaku (Mitsuda \etal~2007)
on 2007 June 7. \suzaku carries four X-ray Imaging Spectrometers (XIS -- Koyama \etal~2007) and
a collimated Hard X-ray Detector (HXD -- Takahashi \etal~2007).
Each XIS consists of four CCD detectors
at the focal plane of its own thin-foil X-ray telescope
(XRT -- Serlemitsos \etal~2007), and has a field-of-view (FOV) of $17.8' \times 17.8'$.
One of the XIS detectors (XIS1) is
back-side illuminated (BI) and the other three (XIS0, XIS2, and XIS3) are
front-side illuminated (FI). The bandpass of the FI detectors is
$\sim 0.4-12$~keV and $\sim 0.2-12$~keV for the BI detector. The useful
bandpass depends on the signal-to-noise ratio of the background-subtracted source data since
the effective area is significantly diminished at the extreme ends of the
operational bandpasses. Although the BI CCD has higher effective area
at low energies, the background level across the entire bandpass is higher
compared to the FI CCDs. The HXD consists of
two non-imaging instruments (the PIN and GSO -- see Takahashi \etal~2007)
with a combined bandpass of $\sim 10-600$~keV.
Both of the HXD instruments are background-limited,
more so the GSO, which has a
smaller effective area than the PIN. 
For AGNs, the source
count rate is typically much less than the background and in the
present study we used only the PIN data, as the GSO data did not provide
a reliable spectrum.
In order to obtain a background-subtracted spectrum,
the background spectrum must be modeled as a function of energy and time.
The background model for the HXD/PIN has an advertised 
systematic uncertainty of 1.3 per cent
\footnote{http://heasarc.gsfc.nasa.gov/docs/suzaku/analysis/pinbgd.html}.
However, the signal is background-dominated, and the
source count rate may be a small fraction of the background count rate,
so the net systematic error in the background-subtracted spectra could
be significant. The observation of Fairall~9 was optimized for the XIS
in terms of positioning the source at the aim point for the XIS (the so-called
``XIS-nominal pointing''), at the expense of $\sim 10$ per cent lower HXD effective area.

The principal data selection and screening criteria for the XIS
were the selection of only {\it ASCA} grades
0, 2, 3, 4, and 6, the removal of
flickering pixels with the
FTOOL {\tt cleansis}, and exclusion
of data taken during satellite passages through the South Atlantic Anomaly
(SAA), as well as for time intervals less than $256$~s after passages through the SAA,
using the T\_SAA\_HXD house-keeping parameter.
Data were also rejected for Earth elevation angles (ELV) less than $5^{\circ}$,
Earth day-time elevation angles (DYE\_ELV) less than $20^{\circ}$, and values
of the magnetic cut-off rigidity (COR) less than 6 ${\rm GeV}/c^{2}$.
Residual uncertainties in the XIS energy scale
are of the order of 0.2 per cent or less
(or $\sim 13$~eV at 6.4~keV -- see Koyama \etal~2007). 
The cleaning and data selection resulted in net exposure times
that are reported in \tablectratesp.

The XIS2 detector was not operational by the time of the Fairall~9 observation
so only data from XIS0, XIS1, and XIS3 were available for analysis.
We extracted XIS spectra of Fairall~9 by using
a circular extraction region with a radius of 3.5'.
Background XIS spectra
were made from off-source areas of the detector, after removing
a circular region with a radius of 4.5' centered on the source,
and the calibration sources (using rectangular masks). 
The background subtraction method for the HXD/PIN used the file 
{\tt ae702043010\_hxd\_pinbgd.evt}, corresponding to the ``tuned'' version
of the background model\footnote{See http://heasarc.gsfc.nasa.gov/docs/suzaku/analysis/pinbgd.html}.

Spectral response matrix
(``RMF'') files, and telescope effective area (``ARF'') files
for the XIS data were made using the \suzaku FTOOLS XISRMFGEN and
XISSIMARFGEN respectively.
The XIS spectra from XIS0, XIS1, and XIS3 were combined
into a single spectrum for spectral fitting. 
The three ``RMF'' and three ``ARF'' files
were all combined, using the appropriate weighting (according to
the count rates and exposure times for each XIS), into a single response file for
the combined XIS background-subtracted spectrum of Fairall~9.
For the HXD/PIN spectrum, the supplied spectral response matrix appropriate for the
time of the observation ({\tt ae\_hxd\_pinxinome3\_20080129.rsp}) was used for spectral fitting.

The useful energy bandpass for the spectrum from each instrument
was determined by the systematics of the background subtraction. 
For the XIS, a useful bandpass of 0.50--9.1~keV was determined, for which the
background was less than 20 per cent of the source counts in any 
spectral bin. The width of the bins for the XIS data was 30~eV.
With this spectral binning, all bins in the 0.50--9.10~keV range
had greater than 20 counts, enabling the use of the $\chi^{2}$ statistic for
spectral fitting. (We avoid grouping spectral bins using a signal-to-noise ratio
threshold because it can ``wash out'' absorption features.)
In addition to omitting data on the basis
of background-subtraction systematics, we also
omitted some spectral data that are subject to calibration uncertainties in
the effective area due to certain atomic features. It is known that
the effective area calibration is poor in the ranges $\sim 1.8$--$1.9$~keV 
(due to Si in the detector) and
$\sim 2.0$--$2.4$~keV (Au M edges due to the telescope). The effective
area also has a significant, steep change at $\sim 1.56$~keV 
(due to Al in the telescope). We took the conservative approach of omitting 
XIS data in the 1.7--1.9 and 2.1--2.35~keV energy ranges 
for the purpose of spectral fitting. For the HXD/PIN, spectral bins with negative
background-subtracted counts clearly indicate a breakdown of the background
model in those bins. We found that the spectral data in the energy range 12.5--42.5~keV 
produced greater than 20 counts per bin (after background subtraction),
for bin widths of 3~keV. This gives the most reliable HXD/PIN spectrum 
with minimal binning whilst at the same time qualifying the spectrum for
spectral fitting using the $\chi^{2}$ statistic.
The selected energy ranges for the XIS and HXD/PIN are
summarized in \tablectratesp, along with the corresponding count rates.

The calibration of the relative cross-normalizations of the XIS and PIN
data involves many factors (e.g. see Yaqoob (2012) for a detailed discussion). 
For observations that had the ``XIS-nominal'' pointing, such as
the present \suzaku observation of Fairall~9, the PIN:XIS ratio
(hereafter $C_{\rm PIN:XIS}$), recommended by the \suzaku Guest Observer Facility (GOF)
is $1.16$\footnote{ftp://legacy.gsfc.nasa.gov/suzaku/doc/xrt/suzakumemo-2008-06.pdf}. 
However, this value does not take into account background-subtraction
systematics, sensitivity to the spectral shape, and other factors that could
affect the actual ratio. On the other hand, allowing $C_{\rm PIN:XIS}$ to be a 
free parameter could potentially skew the best-fitting parameters of a model 
at the expense of settling on a value of $C_{\rm PIN:XIS}$ that is unrelated
to the actual normalization ratio of the two instruments. Consequently, we
performed spectral fits for both cases ($C_{\rm PIN:XIS}$ free, and $C_{\rm PIN:XIS}$
fixed at $1.16$). In addition, a spectral fit was performed using only the XIS data.

We note that Fairall~9 was observed by \suzaku again on 2010 May 19, for $\sim 220$~ks. However,
this observation suffered from severe attitude problems, which have been described in detail
in Lohfink \etal (2012a). The analysis of these data is complicated because the effective
area and spectral response function calculations are compromised. Although Lohfink \etal (2012a)
proceeded to analyse the data by creating a new attitude solution (as described in Nowak \etal 2011),
the software tool that generates the effective area for spectral response functions, XISSIMARFGEN,
was not designed to handle such large excursions in attitude (see Ishisaki \etal 2007 for
a detailed description of XISSIMARFGEN). The resulting errors
in the effective area are unquantified. In fact Lohfink \etal (2012a) pointed out that the XIS 
effective area and spectral 
response function below $\sim 2$ and above $\sim 8$~keV were clearly erroneous. However, the errors
in the effective area in the rest of the data remain unquantified. We therefore did not analyse the
2010 \suzaku data because the results would be unreliable.

\begin{table}
\caption[Exposure times and count rates for the F9 Suzaku spectra]
{Exposure times and count rates for the \suzaku spectra}
\begin{center}
\begin{tabular}{lcccc}
\hline

Detector & Exposure & Energy Range & Rate$^{a}$ & Percentage of \\
         &  (ks)   & (keV)    &  (count $\rm s^{-1}$) & on-source rate$^{b}$ \\
\hline
& & &  \\
XIS & 167.8 & 0.5--1.7, 1.9--2.1, 2.35--9.1 & $0.7535 \pm 0.0013$ & $98.3$ per cent  \\
PIN & 127.3 & 12.5--42.5 & $0.0862\pm0.0021$ & $16.4$ per cent \\
& & &  \\
\hline
\end{tabular}
\end{center}
$^{a}$ Background-subtracted count rate in the energy bands specified. For the XIS, this is the
rate per XIS unit, averaged over XIS0, XIS1, and XIS3.
$^{b}$ The background-subtracted source count rate as a percentage of the total on-source count rate,
in the energy intervals shown.
\end{table}

\section{Modeling the \fekalfa Line and Reflection Continuum}
\label{analysisstrategy}

\begin{figure}
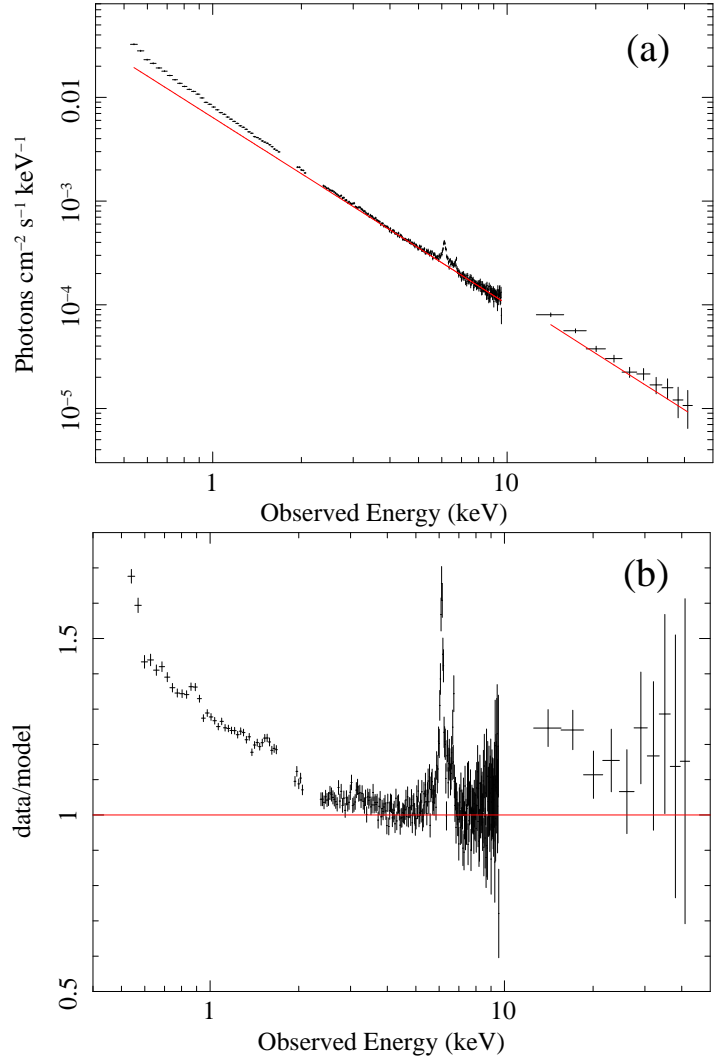

\centerline{
        \psfig{figure=f1a.ps,height=7cm,angle=270}
        }
\centerline{
        \psfig{figure=f1b.ps,height=7cm,angle=270}
        }
        \caption{\footnotesize (a) The unfolded \suzaku XIS and PIN spectra of \srcnamep,
compared to a power-law with a photon index of 1.8. The data below and above 10 keV are
from the XIS and PIN respectively, and the data from XIS0, XIS1, and XIS3
are combined. Note that the PIN:XIS normalization ratio adopted for the plot was 1.16.
(b) The data/model ratio corresponding to (a), in which the \fekalfa line and
high-energy ``hump'' are prominent features, which we show in \S\ref{spfitting} can be 
explained completely by nonrelativistic physics in distant matter, thousands of gravitational
radii from the central black hole.
}
\label{fig:plonlyufspec}
\end{figure}

In Fig.~\ref{fig:plonlyufspec}(a) we show the unfolded \suzaku XIS and PIN spectra of \srcname compared
to a simple power-law continuum with an arbitrary normalization, and photon index, $\Gamma$, of 1.8 
(a value typical of the intrinsic continuum of 
Seyfert galaxies in the pertinent energy range). The actual values of the normalization and
$\Gamma$ are not important here because the purpose of the plot is simply to show the 
salient characteristics of the overall spectrum. Full spectral-fitting will then yield the actual
parameters of the intrinsic continuum. Fig.~\ref{fig:plonlyufspec}(b) shows the
data/model ratio. It can be seen from Fig.~\ref{fig:plonlyufspec} that
above $\sim 3$~keV the spectrum shows the classic form of an \fekalfa emission line 
peaking in the $\sim 6$--$7$~keV rest-frame energy range, with a ``tail'' on the red side of the peak, and
a broad, high-energy continuum ``bump'' above $\sim 10$~keV. In \srcname and many
other AGN, it is these features that
have been interpreted in terms of Fe fluorescence and Compton scattering in the
strong gravity regime in the inner accretion disc within a few gravitational radii of
a spinning black hole (e.g. Walton \etal 2013, and references therein).
Fig.~\ref{fig:plonlyufspec} also shows that the spectrum below $\sim 3$~keV
steepens considerably towards low energies, resulting in a ``soft excess.''

In addition to applying the so-called ``blurred reflection'' model to 
account for the apparent relativistically smeared accretion-disc X-ray spectral signatures 
in the \srcname \suzaku data, Walton \etal (2013) also included
an \adhoc Gaussian component and an additional disc-reflection continuum to model the 
narrow \fekalfa line from distant matter (at tens of thousands of gravitational radii 
from the putative central black hole).
The primary goal of the present spectral-fitting analysis is to determine
whether the \fekalfa emission line and X-ray reflection continuum in the \suzaku X-ray spectrum
of \srcname
can be modeled {\it only} with a narrow \fekalfa emission line and X-ray reflection
continuum from distant matter with a finite column density, with
the line and reflection continuum calculated self-consistently.
Since we will use a model of the X-ray reprocessor that has a finite column density,
we will be able to derive 
constraints on the global column density of the matter that produces the
\fekalfa emission line and X-ray reflection
continuum. Fig.~\ref{fig:plonlyufspec} shows that since the X-ray spectrum of \srcname
is rising continuously towards lower energies, the material that produces the
\fekalfa line cannot be obscuring the line-of-sight. A toroidal geometry for
the X-ray reprocessor, observed at an inclination angle that does not intercept the line-of-sight, 
is therefore consistent with this scenario. (A disc geometry is also compatible, and
has already been applied by Walton \etal 2013, but there is no such model with a finite column
density, suitable for spectral fitting, that is publicly available.)
We will apply the toroidal X-ray reprocessor model, \mytorus (as described in Murphy \& Yaqoob 2009, and
Yaqoob 2012). An alternative spectral-fitting model with a different toroidal geometry to that of the
\mytorus model, due to Brightman \& Nandra (2011), is also available. However, a detailed
study by Liu \& Li (2015) has shown that their model suffers from some erroneous calculations
of the reprocessed X-ray continuum and line spectra, so we did not apply this model. 
We also note that the toroidal model of Ikeda \etal (2009) is not publicly available, and
the CTORUS model (Liu \& Li 2014) only became publicly available after the present study
on Fairall 9 was completed.

Preliminary spectral fitting showed
that a second power-law continuum component, in addition to a principal
intrinsic power-law component for the main hard continuum, was sufficient to account for the steep soft X-ray
continuum . In general, there is a variety in the complexity of the soft excess in AGN,
and there is still much debate as to its origin (e.g. see Middleton, Done \& Gierli\'{n}ski 2007,
Sobolewska \& Done 2007). Although the soft part of the model ionised disc-reflection spectrum of the
blurred reflection model can sometimes account for the soft excess, there are
cases when it cannot. In particular, Lohfink \etal (2012a) found that a
continuum component in addition to the blurred reflection spectrum
had to be introduced to account for the soft
excess in one of the \suzaku observations of \srcnamep. 
The blurred reflection model is certainly not a unique interpretation of the soft excess.
For our purpose, the particular parameterization of the soft X-ray continuum does not
affect our model of the narrow \fekalfa line from neutral matter,
and its associated reflection continuum. Therefore, we used a dual power law for the intrinsic continuum, with the
photon indices and normalizations allowed to be free parameters. We refer to the photon
indices of the hard and soft continua as $\Gamma_{h}$ and $\Gamma_{s}$ respectively.
For the magnitude of the soft component, the actual fitting parameter was the ratio of
the normalization of the soft component to the hard component at 1~keV, instead of
the absolute normalization of the soft component. This ratio is denoted by $r_{\rm PL}$.

In addition to the irrelevance of the details of the soft excess model to our
self-consistent modeling of the \fekalfa line and Compton-reflection continuum,
we found that below 1.5~keV the spectra from the three XIS detectors showed discrepancies with respect
to each other that are larger than the statistical errors. These systematics are
as large as $10$ per cent above 0.7~keV, and $\sim 30$ per cent below 0.7~keV. The systematic
differences in the spectra are likely dominated by the limitations in the
calibration of the effect of contaminants on the CCDs, which have a spatial
and temporal dependence. Above 1.5~keV the spectra from all three XIS detectors
are statistically consistent with each other. Given these considerations, we 
performed all spectral fitting (using the combined XIS spectra) above 1.5~keV.
The energy ranges above 1.5~keV used in the fits were as shown in \tablectratesp.

We did not apply a high-energy cut-off in the form of an exponential diminishing
factor that is often used, because such a form is unphysical. The X-ray reprocessor
model that we use instead has a termination energy for the incident continuum,
and different model tables are available for different termination energies (see Murphy \& Yaqoob 2009).
This is a closer approximation to actual X-ray spectra formed by Comptonisation
in a single-temperature plasma, which
are characterized by a power law form up to some energy, followed by a rollover, as
opposed to the continuous change in slope of exponential cut-off models.  
However, the more physically motivated models are not necessarily
statistically superior to the empirical models in the sense that the two
types of model cannot always be distinguished by the data. For example, in a study of 202 X-ray
observations of Cyg~X-1, Wilms \etal (2006) found that only 12 observations
were better described by a Comptonisation model as opposed to an empirical model
with an exponential cut-off. In the case of Fairall~9,
the \suzaku HXD/PIN spectrum for \srcname only has useful data
extending to $\sim 43$~keV (see Fig.~\ref{fig:plonlyufspec}), so
steepening of the high-energy continuum is not easily discernible, especially 
given the fact that the X-ray reprocessor models themselves cause 
their own high-energy steepening
of the observed spectra, due to Compton downscattering in the circumnuclear matter. 
With \bsaxp, the spectrum of \srcname was measured out to $\sim 100$~keV but
the statistical quality of the data was insufficient to determine if the spectrum
begins to cut off below 100~keV (Dadina 2007). More recently,
multi-epoch fitting of \xmm and \nustar spectra with an exponential cut-off 
yielded only a lower limit on the cut-off energy of 933~keV (Lohfink \etal 2016), 
although the actual value is model-dependent. Therefore, in our modeling of the \suzaku data,
the use of a Comptonisation model for the intrinsic continuum is not necessary. However,
it is important to note that whatever the form of the intrinsic continuum, in the model,
that continuum must extend to higher energies than the highest energy of the data.
This is because high-energy photons can be downscattered (many times if the medium is Compton-thick)
to within the range of the data bandpass, contributing to the observed continuum,
as well as to fluorescent line emission.

We used XSPEC (Arnaud 1996) \footnote{http://heasarc.gsfc.nasa.gov/docs/xanadu/xspec/}
for spectral fitting and utilized the $\chi^{2}$ statistic for minimization.
In all of the model fits we included Galactic absorption with a column density of $3.18 \times 10^{20} \ \rm cm^{-2}$,
obtained with the FTOOL nh, based on the LAB survey described in
Kalberla \etal (2005).
We used photoelectric cross sections given by Verner \etal (1996).
Solar element abundances from Anders \& Grevesse (1989) were used throughout.
All astrophysical model parameter values will be given in the rest frame of \srcname
unless otherwise stated. For the sake of brevity,
certain quantities and details pertaining to particular spectral fits will be given
in the tables of results and not repeated again in the text, unless
it is necessary. Specifically, we are referring to the number of free parameters
in a fit, the number of degrees of freedom, and the null hypothesis probability.
Statistical errors given will be given for one-parameter, 90 per cent confidence (corresponding to
a $\Delta \chi^{2}$ criterion of 2.706).

In a given energy band in the observed frame, observed fluxes will be denoted by $F_{\rm obs}$, and
observed luminosities will be denoted by $L_{\rm obs}$.
These quantities are
not corrected for absorption nor Compton scattering in either
the line-of-sight material, or in the circumnuclear material. 
On the other hand, intrinsic luminosities, denoted by $L_{\rm intr}$ will
be corrected for absorption and X-ray reprocessing. The energy band associated
with a particular value of $L_{\rm obs}$ or $L_{\rm intr}$ will refer to the energy range in the
rest frame of the source. We use a standard cosmology of $H_{0} = 70 \
\rm km \ s^{-1} \ Mpc^{-1}$, $\Lambda = 0.73$, $q_{0} = 0$ throughout the paper.

\subsection{The MYTORUS Model}
\label{mytorusmodel}

The toroidal X-ray reprocessor model, \mytorusp, has
been described in detail in Murphy \& Yaqoob (2009) and Yaqoob \& Murphy (2011). The baseline geometry
consists of a torus with a circular cross section, whose diameter
is characterized by the equatorial column density, $N_{\rm H}$.
The torus is composed of neutral matter and is illuminated by a central, isotropic X-ray source, and
the global covering factor of the reprocessor is $0.5$, corresponding to
a solid angle subtended by the structure at the central X-ray source of
$2\pi$ (which in turn corresponds to an opening half-angle of $60^{\circ}$). 
The \mytorus model 
self-consistently calculates the \fekalfa and \fekbeta fluorescent emission-line spectrum
and the effects of absorption and Compton scattering on the X-ray
continuum and line emission.
The element abundances in the \mytorus model 
are those of the solar values of Anders \& Grevesse (1989), 
and the photoelectric absorption cross-sections
are those of Verner \etal (1996). However, currently, none of the element abundances 
can be varied in the \mytorus model.
The practical implementation of the \mytorus model allows
free relative normalizations between different components of
the model in order to accommodate for differences in the actual geometry
pertinent to the source
(compared to the specific model assumptions used in the original calculations), and 
for time delays between direct continuum, Compton-scattered continuum, and fluorescent 
line photons\footnote{See http://mytorus.com/manual/ for details}. 
However, self-consistency between the fluorescent line emission and the
reflection continuum is always maintained.

The so-called zeroth-order continuum component of the model is the direct,
line-of-sight observed continuum. If the orientation of the toroidal symmetry axis
is such that the angle between that axis and the line-of-sight to the observer
(hereafter \thetaobsp) is greater than the opening half-angle, the zeroth-order
continuum is diminished by absorption and Compton-scattering into
directions away from the line-of-sight. The zeroth-order continuum is thus obtained
from the intrinsic continuum by application of a multiplicative table model and is
not affected by the global geometry of the model.
On the other hand, the global Compton-scattered continuum (or reflection spectrum) is implemented as an XSPEC additive
table model. We used the table {\tt mytorus\_scatteredH200\_v00.fits}, which 
corresponds to a power-law incident continuum with a termination energy
of 200~keV, and a photon index ($\Gamma$) in the range 1.4--2.6. The \fekalfa and \fekbeta emission
lines are implemented with another (single) XSPEC additive table model
that is produced from the same self-consistent Monte Carlo calculations
that were used to calculate the corresponding Compton-scattered continuum table
(we used the table {\tt mytl\_V000010nEp000H200\_v00.fits}).
The parameters for each of the three additive tables are the normalization of the
incident power-law continuum ($A_{\rm PL}$), $\Gamma$, \thetaobsp, and redshift ($z$).

We denote the relative normalization between the scattered
continuum and the
direct, intrinsic continuum, by $A_{S}$, which has
a value of 1.0 for the assumed geometry. This value also implies that either the
intrinsic X-ray continuum flux is constant, or, for
a variable intrinsic X-ray continuum, that the X-ray reprocessor is
compact enough for the Compton-scattered flux to respond
to the intrinsic continuum
on timescales much less than the integration time for the spectrum.
Conversely, departures from $A_{S}=1.0$ imply departure of
the covering factor from $0.5$, or time delays between the intrinsic
and Compton-scattered continua, or both.
However, it is important to note that $A_{S}$ is {\it not} simply
related to the covering factor of the X-ray reprocessor because
the detailed shape of the Compton-scattered continuum varies with
covering factor.
Analogously to $A_{S}$, the parameter
$A_{L}$ is the relative normalization of the \fekalfa line
emission, with a value of 1.0 having a similar meaning to that for $A_{S}=1.0$.
In our analysis we will set $A_{L}=A_{S}$ (relaxing this assumption would
imply significant departure from the default model, such as non-solar Fe abundance).
In practice, $A_{S}$ and $A_{L}$ are each implemented in XSPEC by a ``constant''
model component that multiplies the Compton-scattered continuum and fluorescent line
table respectively, which for $A_{L}=A_{S}$ are tied together. All of the 
remaining parameter values in the various \mytorus tables (described above) are 
required to be tied together by definition of the model tables.
The \mytorus model can be applied in a number of ways that have been detailed in 
Yaqoob (2012), LaMassa \etal (2014), and Yaqoob \etal (2015), but the application to \srcname
and other unobscured AGN is particularly simple because 
only lines of sight that do not intercept the torus are relevant. Therefore,
we do not need to include the zeroth-order continuum component.

\subsection{The \fekalfa Line Energy}
\label{fekalineenergy}

In the \mytorus model the centroid energies of the \fekalfa and \fekbeta lines 
emission are not free parameters
since the lines are explicitly modeled as originating in neutral matter.
In fact, in the \mytorus model, the \fekalfa line is explicitly modeled 
as the doublet K$\alpha_{1}$ at 6.404 keV and K$\alpha_{2}$ at 6.391 keV, 
with a branching ratio of 2:1, which results in a weighted mean
centroid energy of $6.400$~keV (see Murphy \& Yaqoob (2009) for details).
The  \fekbeta line is centered at 7.058~keV.
For \srcnamep, Yaqoob \& Padmanabhan (2004) empirically measured a peak 
rest-frame \fekalfa line energy of
$6.373^{+0.254}_{-0.092}$ keV using high-spectral-resolution \chandra HEG data
and a simple Gaussian parameterization of the line shape (the \fekbeta line
was not detected).

In practice, the peaks of the \fekalfa and \fekbeta emission lines
in the \suzaku data
may be offset relative to the baseline model because of instrumental
calibration systematics and/or mild ionisation. The \suzaku data are
sensitive to offsets in the \fekalfa line peak as small as $\sim 10$~eV.
Therefore, in the \mytorus model we allowed the redshift parameter 
associated with the   
\fekalfa and \fekbeta line table to vary independently of the redshift for
all the other model components (which was fixed at the 
cosmological redshift of \srcnamep). After finding the best-fitting redshift for the
line emission, the line redshift was frozen at that value before deriving
statistical errors on the remaining free parameters of the model.
In the tables of spectral-fitting
results that we will present, the redshift offset will be given as the effective
\fekalfa line energy offset, $E_{\rm shift}$, in the observed frame (or $(1+z)E_{\rm shift}$
in the source frame, where $z$ is the cosmological redshift). A positive shift means that
the \fekalfa line centroid energy is higher than the expected 6.400~keV. In principle,
an energy shift should also be applied to the reflection continuum, but this would require
an additional parameter since the shift is not necessarily going to have the same value as that
for the \fekalfa line. However, since the reflection continuum is broad and the
feature around the Fe~K edge is weak compared to the total continuum, the shift
would not impact the key model parameters and the additional shift
parameter would be poorly constrained. Therefore a shift was not applied
to the reflection continuum.

\subsection{The \fekalfa Line Velocity Width}
\label{fekalinewidth}

In the application of the \mytorus model,
the velocity broadening of the \fekalfa and \fekbeta lines is achieved with 
the {\tt gsmooth} convolution model in XSPEC, which
convolves the intrinsic line emission spectrum with a Gaussian that has a width
$\sigma(E) = \sigma_{0} (E/ 6 \ {\rm keV})^{\alpha}$, where $\sigma_{0}$ and $\alpha$ are the
two parameters of the {\tt gsmooth} model. Since the Doppler velocity width  is
$\Delta v \sim c(\sigma(E)/E)$, fixing $\alpha=1$ models a velocity width that is
independent of energy. The parameter $\sigma_{0}$ is then related to the FWHM by
FWHM~$\sim 2.354 (c/6.0) \sigma_{0} \sim 117,700\sigma_{0} \ {\rm km \ s^{-1}}$, if
$\sigma_{0}$ is in keV. Both the \fekalfa and \fekbeta
lines are tied to a single width parameter, $\sigma_{0}$, since both lines
originate in the same atoms. The spectral resolution
of the \suzaku CCD detectors was of the order $7000 \ {\rm km \ s^{-1}}$~FWHM 
at the \fekalfa line peak energy, at the time
of the Fairall~9 observation. Since the width of the fluorescent emission lines
provides an important constraint on the location of the X-ray reprocessor, we will also 
present, in \S\ref{fekaresults}, a re-analysis of
the \chandra HEG (high-energy grating) data, which has a spectral resolution of
$\sim 1860 \ {\rm km \ s^{-1}}$~FWHM at 6.4 keV. The original analysis of the 
\chandra HEG data (Yaqoob \& Padmanabhan 2004) simply utilized a Gaussian parameterization of the \fekalfa line profile,
whereas the presence of a Compton shoulder in the \fekalfa line profile can affect the
inferred velocity width (e.g. see Yaqoob \& Murphy 2011). The relative magnitude of
the Compton shoulder depends on the column density of the line-emitting matter, and we
will use the \mytorus fits to the \suzaku data to guide fitting the \chandra
HEG data, since the \chandra data have a smaller bandpass and lower signal-to-noise 
ratio than the \suzaku data.

\subsection{The \fekalfa Line Flux and Equivalent Width}
\label{fekalineflux}

The \fekalfa line flux is not explicitly
an adjustable parameter because the line is produced self-consistently
in the \mytorus model of the X-ray reprocessor.
However, by isolating the emission-line table of the \mytorus model,
keeping the best-fitting model parameters, we can measure the observed flux of
the \fekalfa line using an energy range that excludes the \fekbeta line.
(The rest-frame flux is obtained by multiplying the observed flux by ($1+z$).)
The equivalent width (EW) of the \fekalfa line was calculated using the
line flux and the measured (total) 
monochromatic continuum model flux at the observed line peak energy.
The EW in the source frame was then obtained by multiplying the observed
EW by $(1+z)$. Note that the \fekalfa line flux and EW include 
both the zeroth-order and the Compton shoulder
components of the \fekalfa line. Also, in the \mytorus model,
the \fekbeta line flux and EW are not independent of the \fekalfa line
parameters because the theoretical value of the \fekbeta to \fekalfa branching ratio is
already factored into the self-consistent Monte Carlo simulations on which
the model is based (e.g. see Murphy \& Yaqoob 2009).

Since the \fekalfa line flux and EW are not explicit parameters,
the statistical errors on them cannot be obtained
in the usual way. However, in the \mytorus fits, the fractional errors
on the parameter $A_{S}$ are used as approximations for the fractional errors on the
\fekalfa line flux and EW. 

\section{Spectral Fitting Results}
\label{spfitting}

\begin{table}
\caption[Spectral-fitting results]
{Spectral-fitting results}
\begin{center}
\begin{tabular}{lccccc}
\hline
& & & & & \\
Parameter & & & & & \\
& & & & & \\
\hline
& & & & & \\
Mission & \suzaku & \suzaku & \suzaku & \suzaku & \chandra \\
Instruments & PIN,XIS & PIN,XIS & PIN,XIS & XIS & HEG \\
Free Parameters & 13 & 12 & 13 & 12 & 6 \\ 
$\chi^{2}$ / degrees of freedom & 248.7/238 & 254.1/239 & 247.4/238 & 244.3/239 & \ldots \\
Reduced $\chi^{2}$ &              1.045 & 1.063     & 1.039 & 1.067 & \ldots \\
Null Probability                & 0.304 & 0.240     & 0.325 & 0.233 & \ldots \\
$C$ statistic & \ldots & \ldots & \ldots & \ldots & 900.6  \\
``Goodness'' of fit & \ldots & \ldots & \ldots & \ldots & $36.9$ per cent \\ 
\thetaobs ($^{\circ}$) & 0 (f) & 0 (f) & 59 (f) & 0 (f) & 0 (f) \\
$C_{\rm PIN:XIS}$ & $1.32_{-0.11}^{+0.11}$ & 1.16 (f) & $1.30^{+0.12}_{-0.11}$ & \ldots & \ldots \\
$N_{\rm H} \rm \ (10^{24} \ cm^{-2})$ & $0.902^{+0.206}_{-0.216}$& $1.603^{+0.366}_{-0.312}$ 
	& $0.795^{+0.169}_{-0.169}$ & $1.000^{+3.000}_{-0.371}$ & $0.902$ (f)  \\
$A_{S}$ & $1.370^{+0.099}_{-0.086}$& $1.491^{+0.122}_{-0.122}$ & $1.763^{+0.112}_{-0.124}$ 
	& $1.364^{+0.635}_{-0.104}$ & $1.60^{+2.29}_{-1.15}$ \\
$\Gamma_{h}$ & $1.904^{+0.016}_{-0.017}$ & $1.895^{+0.016}_{-0.020}$ & $1.920^{+0.014}_{-0.018}$ 
	& $1.920^{+0.018}_{-0.024}$ & $1.904$ (f)  \\
$\Gamma_{s}$ & $7.10^{+1.67}_{-1.32}$ & $6.88^{+1.54}_{-1.25}$ & $7.60^{+1.73}_{-1.24}$ & 
	$7.10^{+0.16}_{-0.14}$ & $7.10$ (f) \\
$r_{\rm PL}$ & $1.63^{+1.74}_{-0.68}$ & $1.56^{+1.46}_{-0.62}$ & $1.88^{+2.19}_{-0.87}$ 
	& $1.66^{+0.12}_{-0.12}$ & $1.63$ (f) \\
$E_{\rm shift}$ (eV) & $20.0^{+7.9}_{-6.0}$ & $22.2^{+6.5}_{-5.8}$ & $19.7^{+8.1}_{-5.9}$ & 
	$22.2^{+5.8}_{-8.1}$ & $22.5^{+19.6}_{-16.2}$ \\
$I_{\rm Fe~K\alpha}$ ($\rm 10^{-5} \ photons \ cm^{-2} \ s^{-1}$)  & $3.14^{+0.23}_{-0.20}$ & $3.17^{+0.26}_{-0.26}$ 
	& $2.96^{+0.19}_{-0.21}$ & $3.14^{+1.46}_{-0.24}$ & $3.26^{+1.43}_{-0.92}$ \\ 
EW$_{\rm Fe~K\alpha}$ (eV) & $123^{+9}_{-8}$ & $124^{+10}_{-10}$ & $121^{+8}_{-9}$ 
	& $123^{+57}_{-10}$ & $143^{+63}_{-40}$ \\
FWHM[\fekalfap, \fekbeta] ($\rm km \ s^{-1}$)  & $4475^{+1730}_{-2065}$ & $4355^{+1810}_{-2105}$ 
	& $4285^{+1650}_{-2390}$ & $4355^{+1805}_{-2005}$ & $3390^{+11350}_{-1040}$ \\
$E_{\rm Fe~{\sc xxv}}$ (keV)  & $6.752^{+0.038}_{-0.038}$ & $6.744^{+0.048}_{-0.031}$ & $6.751^{+0.042}_{-0.039}$ 
	& $6.752^{+0.037}_{-0.039}$ & $6.752$ (f) \\
$I_{\rm Fe~{\sc xxv}}$ ($\rm 10^{-5} \ photons \ cm^{-2} \ s^{-1}$)  & $0.43^{+0.15}_{-0.15}$ & $0.42^{+0.15}_{-0.15}$ 
	& $0.40^{+0.15}_{-0.16}$ & $0.43^{+0.15}_{-0.15}$ & $<0.82$ \\
EW$_{\rm Fe~{\sc xxv}}$ (eV)  & $16.8^{+5.9}_{-5.9}$ & $17.7^{+6.4}_{-6.5}$ & $16.9^{+6.3}_{-6.8}$ 
	& $18.2^{+6.4}_{-6.4}$ & $<39$ \\
FWHM[Fe~{\sc xxv}] ($\rm km \ s^{-1}$) & 100 (f) & 100 (f) & 100 (f) & 100 (f) & 100 (f) \\
$E_{\rm Fe~{\sc xxvi}}$ (keV)  & $6.970^{+0.039}_{-0.032}$ & $6.975^{+0.033}_{-0.044}$ & $6.975^{+0.035}_{-0.040}$ 
	& $6.975^{+0.034}_{-0.037}$ & $6.970$ (f) \\
$I_{\rm Fe~{\sc xxvi}}$ ($\rm 10^{-5} \ photons \ cm^{-2} \ s^{-1}$)  & $0.41^{+0.16}_{-0.16}$ & $0.39^{+0.16}_{-0.16}$ 
	& $0.38^{+0.16}_{-0.16}$ & $0.40^{+0.15}_{-0.15}$ & $<1.15$ \\
EW$_{\rm Fe~{\sc xxvi}}$ (eV)  & $18.5^{+7.2}_{-7.2}$ & $17.5^{+7.1}_{-7.0}$ & $19.7^{+8.1}_{-5.9}$ 
	& $18.2^{+6.8}_{-6.8}$ & $<58$ \\
FWHM[Fe~{\sc xxvi}] ($\rm km \ s^{-1}$) & 100 (f) & 100 (f) & 100 (f) & 100 (f) & 100 (f) \\
$F_{\rm obs}$[2--10  keV] ($10^{-11} \rm \ erg \ cm^{-2} \ s^{-1}$) & 2.31 & 2.32 & 2.35 & 2.35 & 2.11 \\
$F_{\rm obs}$[10--30  keV] ($10^{-11} \rm \ erg \ cm^{-2} \ s^{-1}$) & 2.09 & 2.33 & 2.12 & \ldots & \ldots \\
$L_{\rm obs}$[2--10 keV] ($10^{44} \rm \ erg \ s^{-1}$) & 1.20 &  1.20 & $1.22$ & 1.22 & 1.09 \\ 
$L_{\rm obs}$[10--30 keV] ($10^{44} \rm \ erg \ s^{-1}$) & 1.08 & 1.20 & $1.10$ & \ldots & \ldots \\
$L_{\rm intr}$[2--10 keV] ($10^{44} \rm \ erg \ s^{-1}$) & 1.22 & 1.19 & $1.15$ & 1.15 & 1.02 \\
$L_{\rm intr}$[10--30 keV] ($10^{44} \rm \ erg \ s^{-1}$) & 0.89 & 0.90 & $0.87$ & \ldots & \ldots \\
\hline
\end{tabular}
\end{center} 
Spectral-fitting results for \suzaku and \chandra HEG data for \srcnamep, obtained from fitting with the \mytorus model.
Fixed parameters are indicated by (f). 
The best-fitting energy shifts of the \fekalfa line model, $E_{\rm shift}$, are given at the line peak in the observed frame.
All remaining parameters are given in the AGN frame. 
In order to obtain the statistical errors on the parameters, in some cases, one or more of the other
parameters had to be temporarily frozen to avoid the spectral fit becoming unstable.
See text for details on this and other aspects of the spectral fitting.
\end{table}

In this section we give the results of fitting the \srcname \suzaku 
data with the \mytorus model, with several variations in the choices of 
parameters that are not allowed to vary. In the \suzaku spectral fits it was also
necessary to include two Gaussian components to model the narrow emission lines
\fexxvres and \feklyap. These lines
originate in an ionised region that must clearly
be distinct from the matter distribution that produces the neutral \fekalfa
line and the associated reflection continuum. They were also included
in previous analyses of the same \suzaku data for Fairall~9 (Schmoll \etal 2009;
Patrick \etal 2011a; Lohfink \etal 2012a; Walton \etal 2013).
The widths of both
Gaussian components were fixed at $100 \rm \ km \ s^{-1}$~FWHM since the
lines are unresolved. Each Gaussian component then has two free parameters,
namely the centroid energy, and the line flux. On the other hand, the width
of the \fekalfa line was allowed to float, and we also give the results of
a re-analysis of a \chandra high-energy grating observation of Fairall~9 in order to
supplement the \suzaku constraints on the \fekalfa line width.
For the sake of reproducibility, we give the complete XSPEC model expression:

\begin{eqnarray}
\rm
\mytorus \ \ model \  = \  constant<1>*phabs<2>( & & \nonumber \\ 
\rm constant<3>*zpowerlw<4> + zpowerlw<5> & + & \nonumber \\
\rm constant<6>*atable\{mytorus\_scatteredH200\_v00.fits\}<7> & + & \nonumber \\
\rm constant<8>*(gsmooth<9>(atable\{mytl\_V000010nEp000H200\_v00.fits\}<10>)) & + &  \nonumber \\
\rm zgauss<11> + zgauss<12> ) \nonumber 
\end{eqnarray}

The model parameters have been described in \S\ref{mytorusmodel}. Here we
identify $\rm constant<1>=C_{\rm PIN:XIS}$, $\rm phabs<2>=$ Galactic column density, ${\rm constant<6>}=A_{S}$,
and ${\rm constant<8>}=A_{L}$. According to the way that the \mytorus model tables are constructed,
the column densities associated with each of the two tables
(components 7 and 10 above) must be tied together, as is also the case 
for the inclination angles. The two intrinsic power-law continuum components are
represented by $\rm zpowerlw<4>$ and $\rm zpowerlw<5>$, associated with photon indices
$\Gamma_{s}$ and $\Gamma_{h}$ respectively. The \mytorus model is linked to only
one of these intrinsic continuum components (the harder power law, with photon index $\Gamma_{h}$).
The component that models the soft X-ray continuum is too steep to
produce any significant \fekalfa line emission and X-reflection continuum, in comparison
to the fluorescence and reflection features due to the flatter, hard X-ray continuum.
The ratio of normalizations of the soft and hard power-law continua 
(i.e. the ratio of fluxes at 1 keV), is denoted by the parameter $r_{\rm PL}$ 
($\rm constant<3>$ in the above model expression).

For lines of sight to the observer that do not intersect the torus (i.e. for $\theta_{\rm obs} <60^{\circ}$),
the shape of the X-ray reflection continuum and the EW of the \fekalfa line are
not very sensitive to the actual inclination angle.
For our baseline spectral-fitting model, we fixed the inclination angle of the torus, $\theta_{\rm obs}$, at $0^{\circ}$,
and allowed $C_{\rm PIN:XIS}$, the \suzaku PIN:XIS normalization ratio, to be free.
The model gives an excellent fit to the data, with a reduced $\chi^{2}$ of 1.045,
and the results are shown in column 2 of \tablemytfitsp.

\begin{figure}
\centerline{
 \psfig{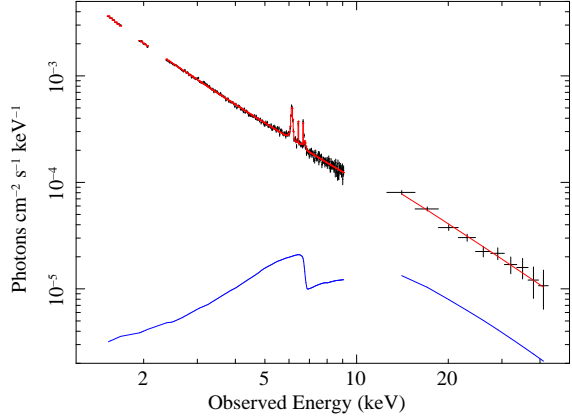}
}
\caption{\footnotesize Broadband spectral fit to the \suzaku \srcname data with the best-fitting face-on \mytorus model
with $C_{\rm PIN:XIS}$ free (see \S\ref{spfitting} and \tablemytfitsp, column 2).
Shown are the unfolded spectral data (black), the net model spectrum overlaid on the data (red), 
and the Compton-scattered (reflection) continuum
contribution (lower, blue curve) that is included in the net model spectrum. Note that in unfolded spectra,
some emission-line features appear to be artificially narrower than they really are.
See Fig.~\ref{fig:fullbanddatzoom}, which shows the same fit as shown here, but 
with the model overlaid on the counts spectrum, which does 
not suffer from this effect. (A color version of this figure is available in the online journal.) 
}
\label{fig:mytfitsufspec}
\end{figure}
 
\begin{figure}
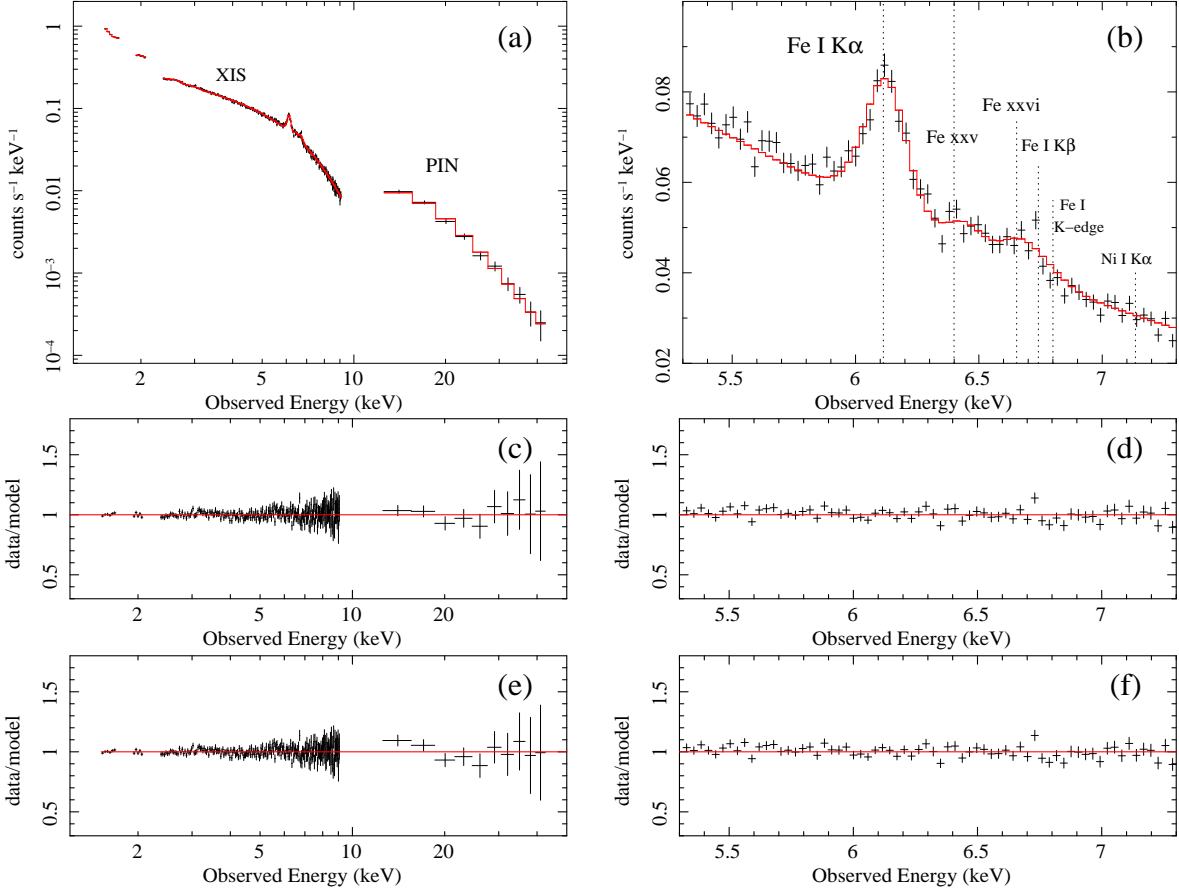

\centerline{
        \psfig{figure=f3a.ps,width=8cm,angle=270}
        \psfig{figure=f3b.ps,width=8cm,angle=270}
        }
\centerline{
        \psfig{figure=f3c.ps,width=8cm,angle=270}
        \psfig{figure=f3d.ps,width=8cm,angle=270}
        }
\centerline{
        \psfig{figure=f3e.ps,width=8cm,angle=270}
        \psfig{figure=f3f.ps,width=8cm,angle=270}
        }
        \caption{\footnotesize (a) The same broadband spectral fit as in Fig.~\ref{fig:mytfitsufspec}
(see \S\ref{spfitting} and \tablemytfitsp, column 2), this time
showing the best-fitting face-on \mytorus model with $C_{\rm PIN:XIS}$ free (red) overlaid on the \suzaku counts spectra (black) for \srcnamep.
Note that data in the energy ranges 1.7--1.9, and 2.1--2.35~keV are
omitted due to calibration uncertainties (see text for details). (b) The data and
model shown in (a), zoomed in on the Fe~K region, showing the detailed fit to the \fekalfa emission line and other atomic features as labeled. Note that the flux of the model \fekalfa line is not controlled by
an arbitrary parameter but calculated self-consistently from the same matter distribution that
produces the global Compton-scattered continuum (see Fig.~\ref{fig:mytfitsufspec}). The dotted lines
correspond to the expected energies of the labeled atomic features in the observed frame. The data to model
ratios corresponding to (a) and (b) are shown in (c) and (d) respectively.
Panels (e) and (f) show the data to model ratios corresponding to the best-fitting face-on \mytorus model with $C_{\rm PIN:XIS}$ fixed at the value of 1.16 (\tablemytfitsp, column 3) for the broadband fit and Fe~K region respectively.
(A color version of this figure is available in the online journal.)
}
\label{fig:fullbanddatzoom}
\end{figure}

This baseline model fit is shown overlaid on the unfolded \suzaku spectrum 
in Fig.~\ref{fig:mytfitsufspec}, with the contribution from the reflection continuum
also shown separately in order to demonstrate its magnitude relative to the total continuum.
Fig.~\ref{fig:fullbanddatzoom}(a) further illustrates the spectral fit, showing the
model overlaid on the counts spectrum, whilst Fig.~\ref{fig:fullbanddatzoom}(b) shows the
data/model ratio. Fig.~\ref{fig:fullbanddatzoom}(c) shows a zoom on the Fe~K band
of the model overlaid on the counts spectrum, and Fig.~\ref{fig:fullbanddatzoom}(d) shows
the corresponding data/model ratio. The data/model ratios show the quality of the
fit, and in particular, the remarkably flat residuals in Fig.~\ref{fig:fullbanddatzoom}(d)
show that the model fits the \fekalfa line, the \fekbeta line, and the Fe~K edge
extremely well. This is despite the fact that no \adhoc or empirical adjustments were applied to
either of the Fe fluorescent lines relative to the Compton-scattered continuum,
and the Fe abundance was fixed at the solar value.
For a given set of model parameters that determine the Compton-scattered
continuum, the Fe fluorescent lines are completely determined. 
This self-consistent solution also shows flat residuals on the
red side of the \fekalfa line peak, implying that no additional complexity in
the model is required. In other words, a relativistically broadened
\fekalfa line is not required, and the narrow \fekalfa line alone is sufficient
to account for the data. Since the data/model ratio in
Fig.~\ref{fig:fullbanddatzoom}(c) is statistically consistent with 1.0, 
it is clear that any additional model complexity would simply be fitting the noise. 
Nevertheless, in \S\ref{relline} we will obtain upper limits on the magnitude
of a contribution from possible relativistically blurred reflection and 
\fekalfa line emission, as well as illustrating how the \mytorus model is able to
reduce the required magnitude of such spectral components.

The best-fitting value of $C_{\rm PIN:XIS}$ for the baseline model is $1.32\pm0.11$ 
(\tablemytfitsp, column 2), which is rather high compared to the
``official'' recommended value of 1.16 for XIS-nominal \suzaku observations. The high
value could be due to background-subtraction systematic uncertainties of the PIN data,
and this effect for the baseline model will be quantified in \S\ref{bgdeffect}.
In order to explore the robustness of the baseline model,
\tablemytfits also shows the results of three variations on the baseline model fit.
In the first of these (column 3 in \tablemytfitsp), the PIN:XIS normalization ratio,
$C_{\rm PIN:XIS}$, is fixed at the value of 1.16.
The data/model ratios for this fit are shown in Fig.~\ref{fig:fullbanddatzoom}(e) and
Fig.~\ref{fig:fullbanddatzoom}(f), and they can be directly compared with the
corresponding ratios for the fit with $C_{\rm PIN:XIS}$ free in Fig.~\ref{fig:fullbanddatzoom}(c)
and Fig.~\ref{fig:fullbanddatzoom}(d) respectively. It can be seen that the
main difference in the data/model ratios for the two fits is that in the fit
with $C_{\rm PIN:XIS}$ fixed there is an excess in the PIN data below $\sim 20$~keV.
On the other hand, in the Fe~K band the differences in the data/model ratios are insignificant 
(compare Fig.~\ref{fig:fullbanddatzoom}(d) and  Fig.~\ref{fig:fullbanddatzoom}(f)).
Next, in a third spectral fit, $C_{\rm PIN:XIS}$ is again allowed to be free, but the inclination angle of
the torus is fixed at $\theta_{\rm obs}=59^{\circ}$ (i.e., just below the value of
$60^{\circ}$ that would make the line of sight graze the surface of the torus).
The results for this case are shown in column 4 of \tablemytfits and can be
used to assess the impact of the different inclination angles on the other model parameters.
In the next fit (column 5 of \tablemytfitsp), the inclination angle is reverted back
to $0^{\circ}$, but this time only the XIS data are fitted. This fit
completely eliminates any systematic uncertainty in the
PIN:XIS normalization ratio,
albeit at the expense of larger statistical errors on the fit parameters.
All of these scenarios give excellent fits to the data, and details of the model parameters obtained from
spectral fitting are discussed below, in \S\ref{mytcontinuum} to \S\ref{bgdeffect}.  
Note that in order to avoid the fits becoming
unstable, some of the parameters had to be fixed in order to derive statistical
errors for the remaining parameters. The parameters that had to
be fixed under such circumstances were the \fekalfa line shift, $E_{\rm shift}$,
the \fekalfa line width, and the centroid energies of the \fexxv and \feklya lines. 
In the fits in which $C_{\rm PIN:XIS}$ was a free parameter, statistical errors
on $C_{\rm PIN:XIS}$ were obtained with the full set of free parameters as 
indicated in \tablemytfitsp. Then statistical errors on $E_{\rm shift}$ were derived,
after which $E_{\rm shift}$, $C_{\rm PIN:XIS}$, 
the \fekalfa line width, and the centroid energies of the \fexxv and \feklya lines were frozen
in order to derive statistical errors on the remaining parameters. The statistical errors on
the \fekalfa line width, and the centroid energies of the \fexxv and \feklya lines were
derived by allowing each parameter to be free in turn.

\begin{figure}
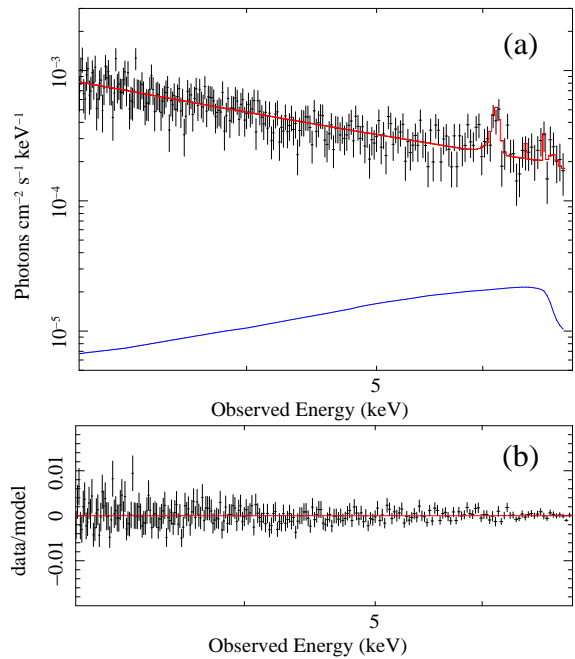

\centerline{
        \psfig{figure=f4a.ps,width=8cm,angle=270}
	}
\centerline{
        \psfig{figure=f4b.ps,width=8cm,angle=270}
        }
        \caption{\footnotesize The baseline \mytorus model fitted to the \chandra HEG 
spectrum (see \S\ref{spfitting} and \tablemytfitsp, column 6), 
showing (a) the best-fitting model overlaid on the unfolded spectrum, and (b) the residuals (difference between data and model counts). (A color version of this figure is available in the online journal.)}
\label{fig:chandrahegfit}
\end{figure}

The final column in \tablemytfits shows the results of fitting the \mytorus model to
\chandra high-energy grating (HEG) data from an observation of Fairall~9 made in 11 September, 2001.
The same spectrum and energy band (2--7~keV) that were used in the original analysis of these
data (as described in Yaqoob \& Padmanabhan 2004) were used to fit the \mytorus model.
We applied the same baseline model that was applied to the \suzaku data (see column 2 of \tablemytfitsp). However, due to
the rather narrow bandpass of the HEG data used, we fixed many of the model
parameters at the best-fitting \suzaku values from the baseline model. The results for
the best-fitting parameters and their statistical uncertainties are shown in column 6 
of \tablemytfitsp. The spectrum and best-fitting model are shown in Fig.~\ref{fig:chandrahegfit}(a), and the residuals
(difference between the data and model counts) are shown in Fig.~\ref{fig:chandrahegfit}(b).
Also indicated in column 6 of \tablemytfits are the parameters that had to be fixed in the
HEG fit. Only six parameters were free, namely the overall continuum normalization, $E_{\rm shift}$, 
$A_{S}$, $\sigma_{0}$ (\fekalfa and \fekbeta line width), and the fluxes of the two
narrow \fexxv and \fexxvi lines from ionised matter. Freezing the other
parameters in the spectral fit does not compromise the principal objective of the
\mytorus fit, which is to attempt to take advantage of the superior spectral resolution of the
\chandra HEG data, in order to supplement the \suzaku constraints on the
velocity width of the \fekalfa line. For the HEG spectral fit we used the 
$C$-statistic for minimization since the counts per channel in the spectrum are
in the Poisson regime. We used the ``goodness'' command in XSPEC to assess the goodness of the fit: 
this command performs Monte Carlo simulations of spectra using the best-fitting model and gives 
the percentage of the simulated spectra that had a fit statistic less than that obtained from the 
fit to the real data. A value of 50 per cent is expected if the best-fitting model is a good representation of the data. 
Values much less than 50 per cent indicate that the data are overparameterized by the model 
since random statistical fluctuations in the majority of the simulated spectra are not able to produce a 
fit statistic as low as that obtained from the real data. In the opposite limit, when 100 per cent of the simulated 
spectra have a fit statistic less than that obtained from the real data, the fit is clearly poor.
Column 6 in Table 2 shows that we obtained a goodness of $36.9$ per cent, indicating that the data
are still somewhat over-parameterized. The results for the width of the \fekalfa line obtained 
from the \chandra HEG fit will be discussed in \S\ref{fekaresults}, along with the \suzaku 
constraints on the \fekalfa line FWHM.

\subsection{Intrinsic Continuum}
\label{mytcontinuum}

In \tablemytfits it can be seen that the photon index of the hard power-law intrinsic continuum,
$\Gamma_{h}$, is very similar for all four \suzaku fits, ranging from $1.895^{+0.016}_{-0.020}$ 
to $1.920^{+0.018}_{-0.024}$. Thus, $\Gamma_{h}$ is not sensitive to the assumed 
inclination angle of the torus, nor to the uncertainty in the PIN:XIS normalization
ratio. The value of $\Gamma_{h} \sim 1.9$ for the 
intrinsic hard X-ray continuum is typical of type~1 AGN
(e.g. Nandra \etal 2007; Dadina 2007 and references therein).
The photon index of the soft continuum component, $\Gamma_{s}$, is much larger (and has
larger fractional statistical errors), but is again similar for all four \suzaku fits,
ranging from $6.88^{+1.54}_{-1.25}$ to $7.60^{+1.73}_{-1.24}$. 
The ratio of the normalization of the soft to hard continuum components, $r_{\rm PL}$,
is also similar for all four \suzaku fits, ranging from $1.56^{+1.46}_{-0.62}$ to
$1.88^{+2.19}_{-0.87}$. Note that for the
\suzaku fit that used only the XIS data (column 5 in \tablemytfitsp), $\Gamma_{h}$ and
$r_{\rm PL}$ had to be fixed at their best-fitting values in order to derive
the statistical errors on $\Gamma_{s}$, otherwise the fit became unstable.
For the same reason, $\Gamma_{s}$ and $\Gamma_{h}$ had to be fixed at their
best-fitting values in order to derive the statistical errors on $r_{\rm PL}$.
This also explains why the statistical errors on $\Gamma_{s}$ and $r_{\rm PL}$
are relatively smaller for the XIS-only fit than for the joint XIS/PIN fits.

Although the values of $\Gamma_{s}$ appear to be unusually high, the soft X-ray
power-law component only makes a significant contribution to the continuum flux
in a narrow bandpass of $\sim 1.5$ to $2$ keV. The largest contribution is at 1.5~keV (the
lowest energy of the fitted range), and even that is only $\sim 11$ per cent of the total
continuum flux. This percentage can be calculated from the values of $r_{\rm PL}$,
$\Gamma_{s}$, and $\Gamma_{h}$ given in \tablemytfitsp. It can also be seen from Fig.~\ref{fig:mytfitsufspec} that
the steepening of the continuum at low energies is barely noticeable. 
The lack of bandpass leverage combined with the weakness of the soft X-ray power law and
small fluctuations due to calibration systematics, conspire
to give a high value of $\Gamma_{s}$. If we extend
the fitted bandpass down to 0.5~keV, smaller values of $\Gamma_{s} \sim 3.5$ are obtained.
However, as described in \S\ref{analysisstrategy}, we did not use data below
1.5~keV for the full spectral analysis because there are larger systematic 
calibration uncertainties below 1.5 keV that manifest as discrepancies between the
three XIS detectors that are larger than the statistical uncertainties in the data.
Since the soft X-ray power-law continuum component is only an empirical parameterization, 
the value of $\Gamma_{s}$ obtained should not be interpreted as having any physical meaning.

\subsection{X-ray Reprocessor Column Densities}
\label{mytcolumns}

\tablemytfits shows that the column
density, $N_{\rm H}$, obtained from the baseline \mytorus model, is 
$0.902^{+0.206}_{-0.216} \times 10^{24} \ \rm cm^{-2}$, which is nearly Compton thick.
Formally, a Compton-thick medium is defined by an optical depth to electron scattering
in the Thomson limit that is greater than unity, or $N_{\rm H} > (1.2 \sigma_{T})^{-1}$, 
or $N_{\rm H} > 1.25 \times 10^{24} \ \rm cm^{-2}$.
The best-fitting value of $A_{S}$ for the baseline fit is $1.370^{+0.099}_{-0.086}$, implying that
the global covering factor of the  X-ray reprocessor may be greater than 50 per cent.
The remaining three \suzaku spectral fits do show significant model dependences
that affect the inferred values of $N_{\rm H}$. In the fit in which \cxispin is 
fixed at $1.16$ (column 3, \tablemytfitsp), $N_{\rm H}$ is $\sim 60$ per cent larger than
the value obtained from the baseline fit (in which \cxispin was free). The 
value of $A_{S}$ is statistically consistent with that obtained from the
baseline fit. In the \suzaku fit in which the torus inclination angle was fixed
at $59^{\circ}$ (column 4, \tablemytfitsp), $N_{\rm H}$ is $\sim 11$ per cent smaller than 
the value obtained from the baseline fit (but the change is
smaller than the statistical errors), whilst $A_{S}$ is $\sim 30$ per cent larger
than the baseline value. Finally, in the \suzaku fit in which
only the XIS data were utilized (column 5, \tablemytfitsp), 
$N_{\rm H} =1.00^{+3.00}_{-0.37}$, which is consistent with the value obtained from the baseline fit.
The corresponding value of $A_{S}$ is also consistent with that from the baseline fit.
In conclusion, the spectral fits 
give a consistent global column density if the \cxispin parameter is allowed to float, or
if only the XIS data are used.
In particular, both the baseline model fit, and the fit
in which only the XIS data are used, are statistically consistent with $N_{\rm H} \sim 
0.9 \times 10^{24} \ \rm cm^{-2}$. Both of these have face-on orientations of the torus, but
other, non-grazing, orientations of the torus have little impact on this inferred column
density. \tablemytfits shows that the nearly grazing inclination angle of $59^{\circ}$ gives
a value of $N_{\rm H}$ that is still statistically consistent with the baseline value.
The derived value of $N_{\rm H}$ is controlled by several aspects of the observed spectrum
working in tandem. The XIS-only fit shows that the lower limit on $N_{\rm H}$ is strongly
controlled by the curvature in the spectrum between $\sim 4$--$10$~keV together with the
flux of the narrow \fekalfa line. The high-energy continuum shape then controls the upper limit
on $N_{\rm H}$ (a value that is too large would produce too much curvature).

\subsection{\fekalfap, \fekbeta Emission Lines, and Fe~K Edge}
\label{fekaresults}

\tablemytfits shows that the best-fitting energy shift of the peak of the \fekalfa line
from the baseline model fit (column 2) is $20.0^{+7.9}_{-6.0}$~eV, and is likely a
result of systematic effects in the XIS energy scale, and possibly mild
ionisation. As far as \fekalfa line fluorescence and Compton-reflection is
concerned, this level of ionisation is negligible and does not conflict with the
assumption that the X-ray reprocessor is essentially neutral. The other three
\suzaku spectral fits give energy shifts consistent with that from the baseline model.
The \chandra HEG fit also gives a consistent energy shift but the statistical errors are
much larger.  

From the baseline \suzaku fit (\tablemytfitsp, column 2), 
the \fekalfa line flux and EW were measured to
be $3.14^{+0.23}_{-0.20} \ \times 10^{-5} \rm \ photons \ cm^{-2} \ s^{-1}$, and
$123^{+9}_{-8}$~eV respectively, where the EW was calculated with respect to 
the total continuum at the line peak (in the AGN frame). The \fekalfa line flux
and EW from the other three \suzaku fits, as well as from the \chandra HEG fit, 
are all consistent with the corresponding values from the baseline \suzaku fit.

The \fekalfa line FWHM from the baseline \suzaku fit is $4475^{+1730}_{-2065} \rm \ km \ s^{-1}$,
and similar values for the FWHM and statistical errors were obtained from the
other three \suzaku fits (see \tablemytfitsp). The \chandra HEG fit gave
FWHM~$=3390^{+11350}_{-1040} \rm \ km \ s^{-1}$. Unfortunately, due to the
limited signal-to-noise ratio of the HEG data, the FWHM has a lower limit that is similar
to the \suzaku lower limit, and an upper limit which is worse than the
upper limit measured from the \suzaku data.
Although the one-parameter, $90$ per cent confidence statistical errors imply that the \fekalfa line is resolved
in all cases, we found that joint, two-parameter confidence contours of the line 
peak energy shift versus line FWHM showed that at 99 per cent confidence the line is not
resolved by either the \suzaku XIS nor the \chandra HEG. Nevertheless, it is clear
that the \fekalfa line could originate from the same location as the optical
broad-line region (BLR) since the \fekalfa line width measurements are consistent with
the widths of some well-studied optical emission lines in Fairall~9. For example,
for H$\beta \ (\lambda 4861)$, FWHM~$=6901\pm 707  \rm \ km \ s^{-1}$ (Santos-Lieo \etal 1997), whilst
for C~{\sc iv}$(\lambda 1549)$ and ${\rm Ly}\alpha \ (\lambda 1216)$, FWHM~$=4628\pm 1375$ 
and $3503 \pm 1474 \rm \ km \ s^{-1}$ respectively (Rodriguez-Pascual \etal 1997).

Using the simple prescription of Netzer (1990) for relating the virialized
velocity of matter orbiting a black hole to the location of the matter, the characteristic radius 
of the line-emitting region can be written $r \sim (4/3)(c/{\rm FWHM})^{2} r_{g}$. Here,
$r_{g} \equiv GM/c^{2} = 1.5 \times 10^{13} \ M_{8} \ \rm cm$ is the gravitational radius,
where $M_{8}$ is the black-hole mass in units of $10^{8} \ M_{\odot}$.
From the \suzaku FWHM measurements in \tablemytfitsp, the largest upper limit
is $6205  \rm \ km \ s^{-1}$ and the smallest lower limit is $1895 \rm \ km \ s^{-1}$
(the \chandra HEG lower limit is higher than this). These values correspond to
a range in ${\rm FWHM}/c$ of $6.32 \times 10^{-3}$ to $2.07 \times 10^{-2}$.
This range corresponds to a spread in the distance of the \fekalfa line emitter from
the central black hole of $3112$ to $33,381$ gravitational radii. This range in
turn corresponds to a distance of $0.015$ to $0.162 \ M_{8}$~pc.
Bentz \& Katz (2015) have compiled a database of AGN black-hole measurements from
BLR optical reverberation mapping campaigns\footnote{http://www.astro.gsu.edu/AGNmass/} that
includes Fairall~9. In addition to measurement errors, there is a large uncertainty
in the mass derived from reverberation campaigns due to the unknown geometry of the
line-emitting region. The compilation for Fairall~9 by Bentz \& Katz (2015;
see also Peterson \etal 2004) yields a range in the black hole mass of 
$M_{8} = 0.99$ to $3.05$. Taking into account this range in the black hole mass, the
distance of the \fekalfa line emitter from the black hole is then $0.015$ to $0.49$~pc.

The \mytorus model self-consistently calculates the Compton shoulder of both
the \fekalfa and \fekbeta lines. A detailed study of the \fekalfa line
Compton shoulder in the \mytorus model has been given in Yaqoob \& Murphy (2011),
and there it is shown that for $N_{\rm H} \sim 10^{24} \ \rm \ cm^{-2}$, and
a face-on torus orientation, the flux in the Compton-shoulder is $\sim 25$ per cent of the
unscattered line core. It can be seen from the model 
overlaid on the data in the Fe~K region in Fig.~\ref{fig:fullbanddatzoom}(b),
and the corresponding remarkably flat residuals in Fig.~\ref{fig:fullbanddatzoom}(d),
that the fit to the entire Fe~K region (including the Compton shoulder)
is excellent. This is the case despite the fact that no empirical adjustments were applied to
either of the Fe fluorescent lines relative to the Compton-scattered continuum, 
and that the Fe abundance was fixed at the solar value. 
For a given set of model parameters that determine the Compton-scattered
continuum, the Fe fluorescent lines are completely determined. 
Not only is the \fekalfa line profile well-fit, but also the \fekbeta line and the Fe~K edge 
are accounted for very well by the model.

\subsection{Fe~XXV and Fe~XXVI Emission Lines}
\label{ionizedlines}

The centroid energies of the (unresolved) Gaussian components used to model the
ionised Fe emission lines were $6.752^{+0.038}_{-0.038}$ and $6.970^{+0.039}_{-0.032}$~keV
from the \suzaku baseline model fit (see \tablemytfitsp, column 2). We identified the
lower and higher energy lines with \fexxvres and \feklya emission respectively. 
The expected energy for the \fexxvres line is $6.700$~keV\footnote{http://physics.nist.gov/PhysRefData},
and that for the \feklya line is $6.966$~keV (Pike \etal 1996).
It can be seen from \tablemytfits that the flux and equivalent width (EW) of the \fexxvres line
from the baseline \suzaku spectral fit
are $0.43^{+0.15}_{-0.15} \times \rm 10^{-5} \ photons \ cm^{-2} \ s^{-1}$ and $16.8^{+5.9}_{-5.9}$ eV
respectively. The flux and EW of the \feklya line, also from the baseline \suzaku fit, are
$0.41^{+0.16}_{-0.16} \times \rm 10^{-5} \ photons \ cm^{-2} \ s^{-1}$ and  $18.5^{+7.2}_{-7.2}$~eV respectively.
The flux and EW of both the \fexxvres and \feklya lines from the other three \suzaku spectral fits
are all statistically consistent with the corresponding values obtained from the \suzaku baseline fit
(\tablemytfitsp). The signal-to-noise ratio of the \chandra HEG data was too poor to detect either
of the two Fe emission lines so only upper limits on the fluxes and EWs are given in \tablemytfitsp.

Narrow emission lines due to \fexxv and \fexxvi from highly ionised matter are common in both
type~1 and type~2 AGN (e.g. see Yaqoob \etal 2003; Bianchi \etal 2005, 2009; Fukazawa \etal 2011;
Patrick \etal 2013; Lobban \& Vaughan 2014). Estimates on the frequency of occurrence of these
emission lines vary. Whilst the frequency of occurrence of these emission lines
has yet to be robustly established, recently, Patrick \etal 2013 estimated the fraction 
of sources with \fexxv and \fexxvi line emission in a sample of 46 type~1 AGN to be 52 and 39 per cent respectively.
The ionised line-emitting matter must be located at a greater distance from the central engine
than the putative torus but an X-ray spectral resolution better than the 
\chandra HEG is required to further tighten the constraints. The \chandra HEG resolution is $\sim 1780$ and $1710 \ \rm km \ s^{-1}$~FWHM at 6.7 and 6.966~keV respectively.

\begin{figure}
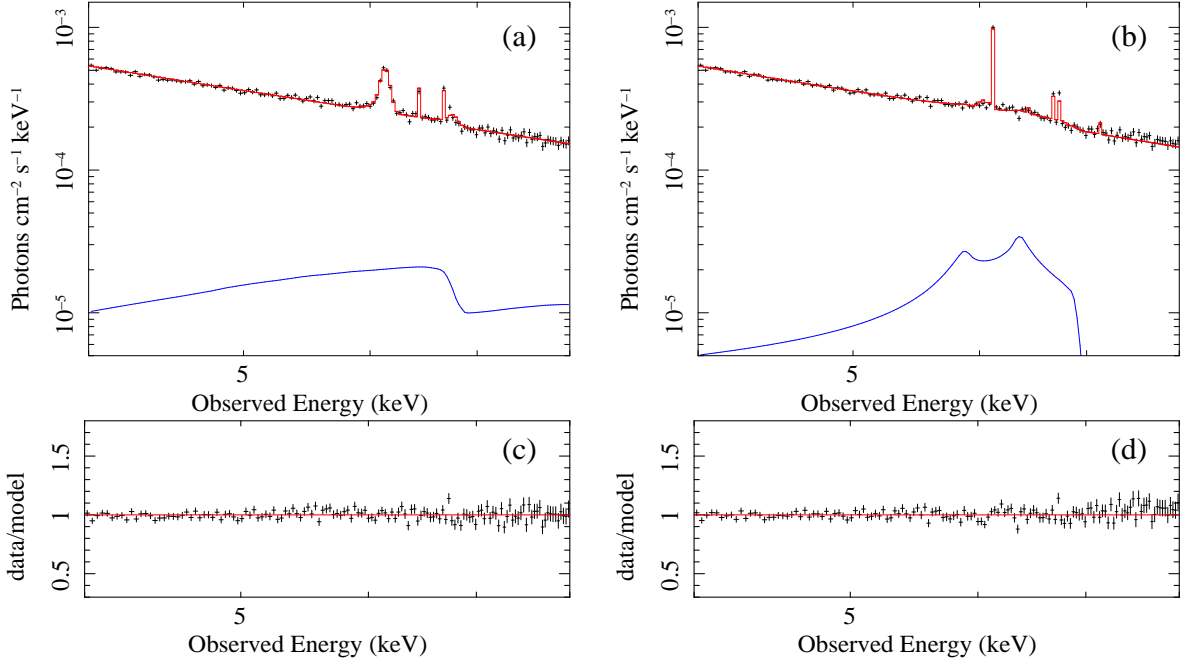

\centerline{
        \psfig{figure=f5a.ps,width=8cm,angle=270}
        \psfig{figure=f5b.ps,width=8cm,angle=270}
        }
\centerline{
        \psfig{figure=f5c.ps,width=8cm,angle=270}
        \psfig{figure=f5d.ps,width=8cm,angle=270}
        }
        \caption{\footnotesize 
A comparison, in the 4--8~keV band, of the best-fitting baseline face-on \mytorus fit with a ``blurred reflection''
model, similar to that fitted type fitted by Lohfink \etal (2012a), showing how the \mytorus model
can eliminate the need for a relativistically broadened \fekalfa line. Both models include narrow \fexxv and \feklya
emission lines from distant ionised material (but the fitted \fexxv flux for the blurred reflection model is
negligible). (a) Unfolded photon spectrum (black) overlaid with the total
\mytorus model (red; see \tablemytfitsp, column 2). The Compton-scattered continuum from the \mytorus model shown
in blue is able to account for excess emission in the Fe~K band instead of a relativistically broadened \fekalfa line.
(b) Unfolded photon spectrum (black) overlaid with the total multi-component Lohfink \etal (2012a) 
disc reflection and Comptonisation model in red (see text for details).
The blurred reflection component is shown in blue. Above the
blue wing of the relativistic \fekalfa line, the reflection continuum drops by more than an order of magnitude but rises again at
high energies. Note that in this model fit, the narrow \fekalfa
is unresolved because Lohfink \etal (2012a) did not include any broadening for this component. 
Panels (c) and (d) show the data to model ratios corresponding to (a) and (b) respectively, showing that both models
give good fits to the data. Note that, although the \mytorus component and blurred reflection component
(shown in blue in panels (a) and (b) respectively) appear to be markedly different
from each other, smearing due to the limited CCD detector energy resolution has a considerable
impact on the ability of the data to distinguish between the two models. This is illustrated in Fig.~\ref{fig:cmprelline}.
(A color version of this figure is available in the online journal.)
}
\label{fig:cmprelfits}
\end{figure}

\subsection{Fluxes and Luminosities}
\label{mytfluxes}

Continuum fluxes and luminosities obtained from all the spectral fits are shown in
\tablemytfits for the 2--10~keV band, and for the 10--30~keV energy bands for the \suzaku fits. For fluxes,
these energy bands are in the observed frame, but for luminosities, the lower and upper energies
in each band are quantities in the source frame, so that they appear redshifted in the 
observed frame. For regions where there were no data, the models were extrapolated.
Two luminosities are given for each band, one set ($L_{\rm obs}$) was derived directly from the 
best-fitting model, and the second set, $L_{\rm intr}$, or intrinsic luminosity, was derived
by removing all absorbing and reflecting model components.

The 2--10~keV observed fluxes are similar for
all four \suzaku spectral fits, lying in the range $2.31$ to $2.35 \times 10^{-11} \ \rm erg \ cm^{-2} \ s^{-1}$.
The chandra HEG 2--10~keV flux is $2.11 \times 10^{-11} \ \rm erg \ cm^{-2} \ s^{-1}$.  
The observed 2--10~keV luminosity is $\sim 1.2 \times 10^{44} \  \rm erg \ s^{-1}$ for the \suzaku fits,
and $\sim 1.1 \times 10^{44} \  \rm erg \ s^{-1}$ for the \chandra HEG fit.
The intrinsic 2--10~keV luminosities are very similar to each other for all of the spectral fits,
since the 2--10~keV band is dominated by the intrinsic continuum.
The 10--30 keV fluxes are in the range $2.09-2.33 \times 10^{-11} \ \rm erg \ cm^{-2} \ s^{-1}$ for
the four \suzaku fits, and the corresponding observed 10--30 keV luminosities are in the range
$1.08$--$1.20 \times 10^{44} \  \rm erg \ s^{-1}$. The intrinsic 10--30~keV luminosities are 
in the range $0.87-0.90 \times 10^{44} \  \rm erg \ s^{-1}$, somewhat less than the observed
luminosities, because the latter include reflection from the torus which increases the 
apparent luminosity. Note that the intrinsic 10--30~keV luminosities were derived using the XIS normalizations,
so they do not include the effect of different values of the PIN:XIS normalization ratio
(which does, however, affect the observed luminosities).

\subsection{HXD/PIN Background Subtraction Systematics}
\label{bgdeffect}

In this section we evaluate the effect of a $\pm 1$ per cent error in the HXD-PIN background subtraction
on the key parameters of the baseline \mytorus \suzaku model fit (\tablemytfitsp, column 2).
Both scenarios did not significantly change the quality of the fit, with $\Delta \chi^{2} <1$ for
both cases. Moreover, the effect on continuum fluxes for both cases was negligible.
Compared to the baseline fit values, the
2--10~keV fluxes and the 10--30~keV fluxes changed by $<2$ per cent.
On the other hand,
we found that changing the  background by $+1$ per cent gave 
$N_{\rm H} = 0.895^{+0.456}_{-0.341} \ \times 10^{24} \ \rm cm^{-2}$,
$\Gamma = 1.905^{+0.019}_{-0.020}$, $A_{S} = 1.377^{+0.170}_{-0.091}$,
and $C_{\rm PIN:XIS} = 1.26^{+0.102}_{-0.105}$.
Changing the background by $-1$ per cent gave 
$N_{\rm H} = 0.894^{+0.338}_{-0.495} \ \times 10^{24} \ \rm cm^{-2}$, 
$\Gamma = 1.905^{+0.018}_{-0.021}$, $A_{S} = 1.375^{+0.167}_{-0.090}$,
and $C_{\rm PIN:XIS} = 1.390^{+0.109}_{-0.111}$.
Thus, compared to the corresponding values with standard background in \tablemytfitsp, column 2,
only the value of $C_{\rm PIN:XIS}$ is impacted and there is no statistically significant
change to the other parameters. In the case that the background is changed by $+1$ per cent,
the value of $C_{\rm PIN:XIS}$ becomes consistent with the ``recommended value'' of 1.16 for
XIS-nominal observations. We note that the background-subtraction systematics cannot be
completely described by a single number and in reality have an
energy dependence. Also, they are largest between $\sim 10$--20~keV, and the residuals as a function of energy,
obtained when the model is applied to different Earth occultation data sets, are not identical\footnote{ftp://legacy.gsfc.nasa.gov/suzaku/doc/hxd/suzakumemo-2008-03.pdf}.
The pattern of the PIN residuals in the fit with $C_{\rm PIN:XIS}$
fixed at 1.16 (Fig.~\ref{fig:fullbanddatzoom}(e)) could be the result of the limitations of
the background model. Thus,
the value of $C_{\rm PIN:XIS}=1.32\pm0.11$ from the baseline fit does not appear to be
too unreasonable.

\subsection{Comparison with Relativistic Line Models}
\label{relline}

In this section we investigate how the Compton-scattered continuum in the \mytorus fits to the
\srcname \suzaku data is able to mimic a relativistically broadened \fekalfa line by directly comparing
the face-on \mytorus fit with a so-called ``blurred reflection'' model fitted by Lohfink \etal (2012a)
to the same data. We do not delve in detail into fitting ionised accretion disc models with relativistic blurring
because such model fits with several variations on the basic theme have been described at length in Lohfink \etal (2012a)
and there is little to gain from repeating the analysis. Instead, we take one of the preferred models
of Lohfink \etal (2012a) and directly compare the shape of the blurred reflection spectrum, which includes
the relativistic \fekalfa line, with the component of the \mytorus model that replaces the relativistic
emission. Specifically, we adopt the blurred reflection model fit shown in Table 4 (under the ``{\it Suzaku A}'' column)
in Lohfink \etal (2012a). This is one of the two preferred models 
from multi-epoch fitting in that work, and has one ionised
reflection component, whereas the other model is more complicated and has two ionised reflection components.
The ionised reflection is modeled with the REFLIONX model, and this is convolved with the model RELCONV
to implement the relativistic blurring. The Lohfink \etal (2012a) fit employs
additional model components, namely PEXMON for the narrow, distant-matter
\fekalfa line and reflection continuum; COMPTT for a Comptonisation continuum; and an XSTAR table
for an ionised plasma component. However, we did not include the latter because it is relatively weak
and does not impact the fitted reflection or other continuum components.
Although the \fexxv and \fexxvi
emission lines are also relatively weak, we modeled them with individual, unresolved Gaussian components,
as was done for the \mytorus fits described in \S\ref{spfitting}, allowing only the centroid energies and
line fluxes to be free parameters. We froze the parameters of the remaining model components at the values
in Table~4 of Lohfink \etal (2012a), including $C_{\rm PIN:XIS}=1.16$, except for the normalizations 
of continuum components, and fitted the model to the same \suzaku XIS and PIN spectra that were used
for fitting the \mytorus model. Specifically, the power-law photon index was $\Gamma=1.90$, the distant-matter
reflection fraction for PEXMON was $0.84$, the disc ionisation parameter was $\xi = 10.1$, the 
disc inclination angle was $48^{\circ}$, the radial emissivity index was $q=2.0$, the
iron abundance in the REFLIONX model relative to solar was 10.0, and the black-hole spin
was $a=0.52$. The temperature of the Comptonisation component was $kT = 20.52$ keV, and the optical depth
of the Comptonising matter was $\tau =0.52$.

In Fig.~\ref{fig:cmprelfits}(a) and Fig.~\ref{fig:cmprelfits}(b) we compare, in the critical 4--8~keV band,
the unfolded spectra for the face-on \mytorus fit and for the blurred reflection model respectively.
The data to model ratios are shown in Fig.~\ref{fig:cmprelfits}(c) and Fig.~\ref{fig:cmprelfits}(d), from
which it can be seen that neither model leaves significant residuals. 
For the blurred reflection model we obtained $\chi^{2} = 368.0$ for 244 degrees of freedom,
corresponding to a reduced $\chi^{2}$ value of 1.508 and null hypothesis probability of $4.8 \times 10^{-7}$.
However, it is principally the residuals at energies below $\sim 3$~keV that are responsible
for the high value of $\chi^{2}$ since we did not include the photoionised plasma component.
We note that Lohfink \etal (2012a) give $\chi^{2}=2232.6$ for 2095 degrees of freedom,
or a reduced $\chi^{2}$ value of 1.066 (null hypothesis probability of $0.02$), 
for the full blurred reflection model that they fitted, 
which did include the photoionised plasma emission. However,
there are additional differences between our blurred reflection model fit and that of Lohfink \etal (2012a).
One is that the latter used two separate XIS spectra (one combining XIS0 and XIS3 data, the other 
corresponding to XIS1 data only). In addition, Lohfink \etal (2012a) employed
a different energy band for fitting the PIN data, namely 16--35~keV, as opposed 12.5--42.5~keV,
and they used XIS data up to 10~keV.
When we repeated our blurred reflection model fit with data below 3 keV excluded, 
 using two separate XIS spectra as described, and the same energy ranges as Lohfink \etal (2012a), we
obtained a reduced $\chi^{2}$ value of 1.099 (null hypothesis probability of $0.065$). 
Considering the fact that our data selection criteria and extraction regions for XIS source and
background data are not completely identical, we conclude that our blurred reflection fit 
results are consistent
with those of Lohfink \etal (2012a). The blurred reflection model fitted by Lohfink \etal (2012a)
and the \mytorus model fit (see \tablemytfitsp) give reduced $\chi^{2}$ values that are comparable,
which indicates that the two model fits are statistically similar to each other.
In our blurred reflection model fit described above, the value of \cxispin was 
fixed at 1.16, as it was in Lohfink \etal (2012a). When we allowed \cxispin to be a free
parameter, the value of $\chi^{2}$ dropped by only 1.6, and we obtained
$C_{\rm PIN:XIS} = 1.10_{-0.08}^{+0.08}$, which encompasses the value of 1.16.

\begin{figure*}
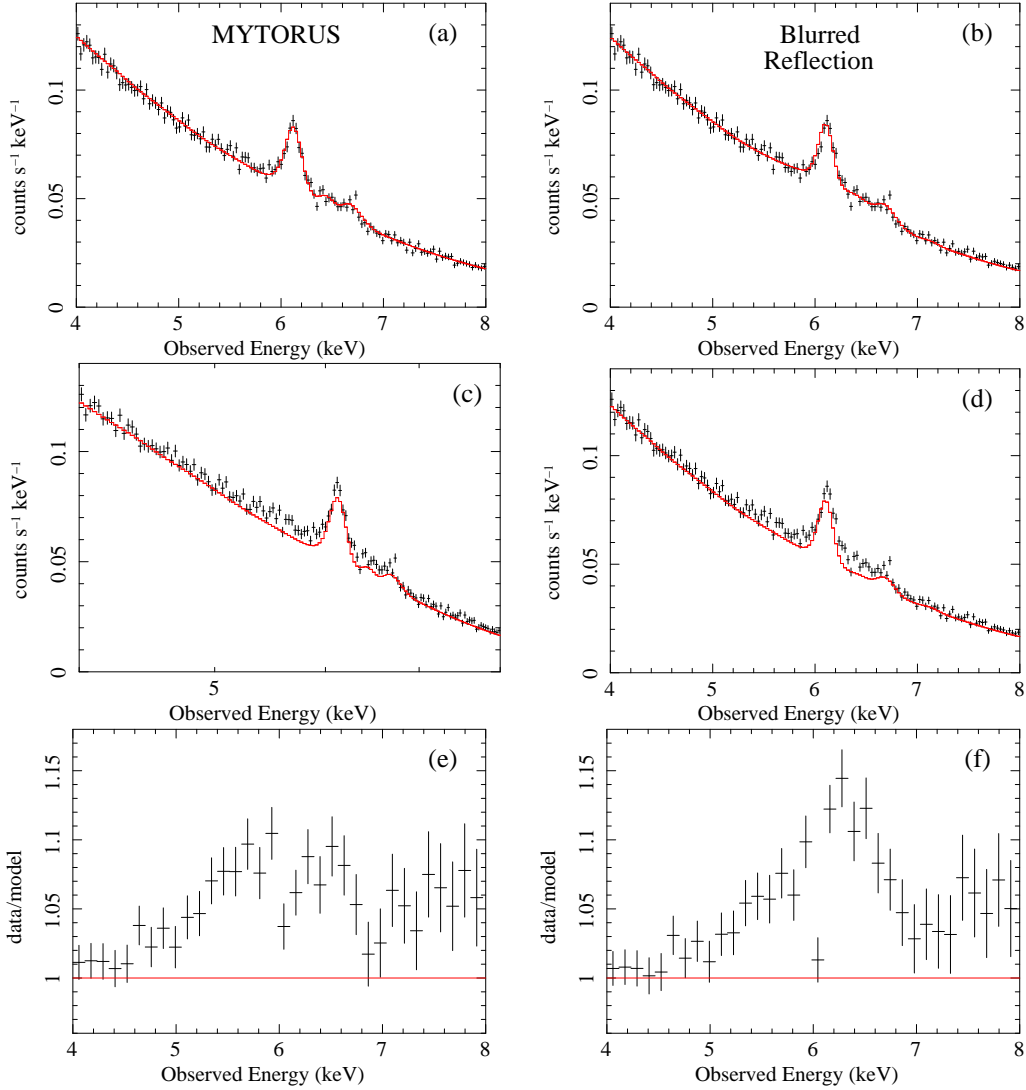

\centerline{
        \psfig{figure=f6a.ps,width=7cm,angle=270}
        \psfig{figure=f6b.ps,width=7cm,angle=270}
        }
\centerline{
        \psfig{figure=f6c.ps,width=7cm,angle=270}
        \psfig{figure=f6d.ps,width=7cm,angle=270}
        }
\centerline{
        \psfig{figure=f6e.ps,width=7cm,angle=270}
        \psfig{figure=f6f.ps,width=7cm,angle=270}
        }
        \caption{\footnotesize  
The effect of turning off the Compton-scattered continuum in the best-fitting baseline face-on \mytorus model, compared
with turning off the reflection component (which includes the relativistically broadened \fekalfa line) in
the blurred reflection model (see text for details). (a) XIS counts spectrum of \srcname (black),
overlaid with the best-fitting face-on \mytorus (red; see \tablemytfitsp, column 2). (b) XIS counts spectrum of \srcname (black),
overlaid with the blurred reflection model, similar to that fitted by Lohfink \etal (2012a),
and described in \S\ref{relline}. (c) As (a) but with the Compton-scattered
continuum in the \mytorus model turned off. (d) As (b) but with the blurred reflection component turned off.
Panels (e) and (f) show data to model ratios corresponding to (c) and (d) respectively. In the blurred reflection fit,
the distant-matter, narrow \fekalfa is unresolved because Lohfink \etal (2012a) did not allow it to be free. It can be
seen that consequently, the relativistic \fekalfa line in the blurred reflection model is actually modeling part of the
the narrow \fekalfa line. Otherwise, the feature in (e) modeled by the Compton-scattered continuum of the \mytorus model is, 
as expected, very similar to the broad feature in (f) modeled as a relativistic line.
Note that for clarity, the data/model ratios are binned by a factor of 4 compared to the spectral plots.
(A color version of this figure is available in the online journal.)
}
\label{fig:cmprelline}
\end{figure*}

The model components in blue 
in Fig.~\ref{fig:cmprelfits}(a) and Fig.~\ref{fig:cmprelfits}(b) show how the Compton-scattered continuum
from the \mytorus model and the relativistically broadened \fekalfa line with blurred disc continuum respectively
provide very different solutions to the same data. At a disc inclination angle of $48^{\circ}$, the
relativistic line has significant emission in the blue wing but it also has a trailing red wing. In the
\mytorus model, the downturn at low energies of the Compton-scattered continuum, combined with the Fe~K edge in 
that continuum conspire to mimic the relativistic line in the blurred reflection model. Note that
in the blurred reflection model the reflection continuum above the blue wing of the broad \fekalfa line
drops by more than order of magnitude but recovers again at higher energies (see plots in Lohfink \etal 2012a).
Obviously, if the data had better spectral resolution, the two models may be distinguishable, but at CCD spectral resolution
they are not distinguishable (this is demonstrated in Fig.~\ref{fig:cmprelline}, discussed below).
It is also important to note that it can be seen
that the Compton-shoulder in the \mytorus model fit
is not the reason why the \mytorus model is able to mimic the relativistic line in \srcnamep, in part
because the shoulder cannot explain the blue wing, but also because with the velocity broadening of the
narrow \fekalfa line of $\sim 4475^{+1730}_{-2065} \ \rm km \ s^{-1}$~FWHM (see \tablemytfitsp), the
Compton shoulder cannot be discerned as a distinct feature. This is expected, especially for
data with CCD spectral resolution (see Yaqoob \& Murphy 2011 for a detailed discussion). 
The velocity broadening of the narrow, distant-matter \fekalfa line highlights another important difference
between the \mytorus and blurred reflection model fits. Lohfink \etal (2012a) did not apply velocity
broadening to the distant-matter \fekalfa line, which is part of the PEXMON model in their blurred reflection
fit. No justification was given for this omission, so it is an implicit assumption of their
model that the line-emitting matter is sufficiently far from the central mass that Doppler broadening
is negligible compared to the CCD energy resolution. The result of this assumption is that the
relativistic line in the blurred reflection model partly models the distant-matter narrow \fekalfa line,
thereby exaggerating the relative importance of the relativistic line. This is illustrated in
Fig.~\ref{fig:cmprelline}: the counts spectra overlaid with the \mytorus and blurred reflection models
are shown in Fig.~\ref{fig:cmprelline}(a) and Fig.~\ref{fig:cmprelline}(b) respectively, which in turn
correspond to the same fits shown against the unfolded spectra in 
Fig.~\ref{fig:cmprelfits}(a) and Fig.~\ref{fig:cmprelfits}(b) respectively. The
result of turning off the Compton-scattered continuum in the \mytorus model is then shown in Fig.~\ref{fig:cmprelline}(c),
whilst the result of turning off the blurred reflection component in the Lohfink \etal (2012a) fit is shown
in Fig.~\ref{fig:cmprelline}(d). In the latter, it can be seen that turning off the relativistic \fekalfa line
in the model leaves an excess in the data on blue side of the distant-matter \fekalfa line. This is not
the case for \mytorus model. The data to model ratios corresponding to Fig.~\ref{fig:cmprelline}(c) and
Fig.~\ref{fig:cmprelline}(d) are shown in Fig.~\ref{fig:cmprelline}(e) and Fig.~\ref{fig:cmprelline}(f) 
respectively (note that the ratios are binned by a factor of 4 for clarity compared to the spectral plots).
The data to model ratios further confirm that the broad, relativistic \fekalfa line is modeling part
of the narrow, distant-matter \fekalfa line, and that aside from this difference,
the feature in Fig.~\ref{fig:cmprelline}(e) modeled by the \mytorus model Compton-scattered continuum
is, as expected, very similar to the broad feature modeled as a relativistic line in Fig.~\ref{fig:cmprelline}(f).
Whilst the limited signal-to-noise ratio of the data plays a role in allowing both
models to fit the data, the smearing due to the limited energy resolution of CCD data 
is a key factor. If the energy resolution were better, the data to model ratios in 
Fig.~\ref{fig:cmprelline}(e) and Fig.~\ref{fig:cmprelline}(f), which by definition
have the models folded through the instrument response function, would reflect more clearly the
differences in spectral shape between the two different models of the broad spectral feature.

Whereas the above analysis considered the \mytorus and blurred reflection models separately,
next we consider the upper limit on a blurred reflection model when it is added to the baseline \mytorus
model (\tablemytfitsp, column 2). Since it is clear from the residuals to the \mytorus model (e.g.,
Fig.~\ref{fig:fullbanddatzoom}(d)) that adding a relativistic \fekalfa line would essentially result
in fitting the noise in the data, the question of an upper limit on a blurred reflection component
must be posed in a very specific context in order to give a meaningful result. Allowing
too many free parameters may result in none of them being constrained. Thus, we 
added a REFLIONX model component, convolved with the RELCONV model, and again
tested against the best-fitting values of the parameters given in Table~4 of Lohfink \etal (2012a),
fixing those parameters in the fits, except for the normalization of the REFLIONX component.
We then obtained the 1-parameter, $90$ per cent confidence upper limit on this normalization.
The upper limit that we obtained corresponds to a 2--10~keV flux for the blurred reflection
component that is only 1.2 per cent of
the total 2--10~keV flux.

In the REFLIONX model, the \fekalfa line from the disc cannot be separated from the 
disc-reflection continuum, so the equivalent width (EW) of the \fekalfa cannot easily be calculated. 
Therefore,  we also considered the upper limit on the EW of a relativistically 
broadened \fekalfa line that is not attached to a reflection continuum, when it
is added to the baseline \mytorus model fit. Inclusion of a reflection continuum would only
reduce the upper limit on the EW. We tested the data against the LAOR model which gives
a line profile due to emission from a disc associated with a maximally spinning
black hole ($a=0.998$; see Laor 1991). We fixed the inner disc radius at a value
appropriate for the spin, namely $1.235$ gravitational radii, and we fixed the outer
disc radius at 400 gravitational radii. We also fixed the radial emissivity index at $2.0$.
The disc inclination angle was fixed at the value of $48^{\circ}$ (i.e. the value in Table 4 of Lohfink \etal 2012a).
We obtained a 1-parameter, 90 per cent upper limit on the EW of 27~eV. 
We repeated the procedure with the inclination angle fixed at $0^{\circ}$ and
again obtained a small upper limit on the EW of the relativistic line of 32~eV.
Using line profiles corresponding to smaller values of black-hole spin would result in even
smaller upper limits on the EW because the innermost stable orbit radius increases with
decreasing black-hole spin, so the corresponding line profiles would have decreasing width.

\section{Summary}
\label{summary}

We have re-analysed \suzaku data for the ``bare'' Seyfert~1 galaxy, \srcnamep, applying
the \mytorus model of Murphy \& Yaqoob (2009) that self-consistently calculates
the \fekalfa line emission and X-ray reflection spectrum from a toroidal
distribution of neutral matter with solar iron abundance. We find that an excellent fit is obtained
for the detailed \fekalfa line profile including the red and blue wings, the \fekbeta line, 
the Fe~K edge, and the ``Compton hump'' associated with the X-ray reflection spectrum.
The same best-fitting parameters that determine
the magnitude and shape of the X-ray reflection continuum also determine the \fekalfa
line flux and profile with no additional freedom. The data require no \adhoc adjustment
or interference of this self-consistency. We obtained an equatorial column density of
the toroidal X-ray reprocessor of $N_{\rm H} = 0.902^{+0.206}_{-0.216} \ \rm \times 10^{24} \ cm^{-2}$
from a baseline fit,
for the case that the torus is observed face-on and the cross-normalization between
the PIN and XIS data is allowed to be free.
Other lines of sight that do not intercept the torus also provide excellent fits to the
data, yielding similar constraints on the column density. A face-on fit with the PIN data
removed altogether also provides consistent constraints on $N_{\rm H}$. On the other hand,
fixing the cross-normalization between the PIN and XIS spectra at the default value
gives a column density that is $\sim 60$ per cent higher than that from the baseline fit.
 
The baseline, face-on toroidal model gives a velocity broadening 
of the \fekalfa line of $4475^{+1730}_{-2065} \ \rm km \ s^{-1}$~FWHM. Variations on the baseline model also
gave consistent values for the  FWHM of the \fekalfa line. The FWHM measured here
does not confuse broadening due to the Compton shoulder of the \fekalfa line with
velocity broadening since the broadening due to the Compton shoulder is self-consistently modeled. 
The  \fekalfa line width places the distance of the
X-ray reprocessor at $\sim 3.1$ to $33.4 \times 10^{3}$ gravitational radii from the putative central black hole.
This corresponds to $\sim 0.015$ to $0.49{\rm \ pc}$ for a black-hole mass in the 
range $0.99-3.05 \times 10^{8} \ M_{\odot}$ (from historical reverberation measurements). 
The statistical errors quoted here for the FWHM are for one parameter, 90 per cent confidence, but we found that
the \fekalfa line is not resolved at a confidence level of 99 per cent, for two parameters.
Thus, a location for the \fekalfa line emitter that is further out than $\sim 0.5$~pc cannot
be ruled out. We also detected narrow, unresolved \fexxvres and \feklya emission lines from highly 
ionised matter in a region distinct from the \fekalfa line emitter, further out from the central engine.

The \mytorus models, using only mundane, nonrelativistic physics,
give such good fits to the \srcname data using only narrow \fekalfa line
emission and X-ray reflection from distant matter that there is no need for
any more complexity in the models. In contrast,
previous analyses of the same \suzaku data for \srcname have applied relativistic
disc-reflection models that produce broad \fekalfa line emission and X-ray reflection from within
a few gravitational radii of the black hole (Lohfink \etal 2012a, and references therein). Such fits have been used to
directly derive constraints on the black-hole spin, or angular momentum.
In our interpretation of the data, there are no signatures in the X-ray spectrum
of the strong gravity regime, no broad \fekalfa line, and therefore no measure of black-hole spin
based on the \fekalfa line and reflection spectrum. 
From a theoretical perspective, there is actually no precedent for emission features
from within a few gravitational radii of the black hole to leave imprints on the X-ray 
spectrum since there are several ways to suppress such features. For example, the
inner accretion disc may be truncated, or it may be too highly ionised (e.g. see
Patrick \etal 2013 for a recent discussion). 
We showed in detail how the spectral shape of the \mytorus model in the Fe~K band
is able to replace the relativistically broadened \fekalfa line, but we also obtained
an upper limit to the magnitude of a relativistically blurred accretion disc-reflection spectrum
if it is included in addition to the \mytorus model. Relative to the baseline \mytorus model,
the 2--10~keV flux in the blurred reflection spectrum is $<1.2$ per cent of the total
observed 2--10~keV flux.

An important factor in our \mytorus spectral fits 
is the break from the common assumption that the X-ray reflector and fluorescent line emitter 
has an infinite column density. The X-ray reflection continuum from matter with a finite column density
has a greater variety of spectral shapes than that from matter with an infinite
column density. In particular, reflection spectra from
matter with a finite column density can produce spectral structure around the \fekalfa line
that might otherwise be interpreted as the effects of relativistic smearing. 
Moreover, we note that in previous analyses of the X-ray spectra of \srcname that derived 
black-hole spin measurements, the infinite column density disc model was not only applied
to what was thought to be the relativistic components, but it was also used to model the
distant-matter (narrow) \fekalfa line and reflection spectrum. In future work we will
apply the finite column density reprocessor models described here to other AGN and other
observations of Fairall~9 in order to investigate whether our conclusions have a broader relevance.

\vspace{5mm}
\noindent
Acknowledgments \\
The authors thank the anonymous referee for helping to improve the paper.
The authors acknowledge support for this work from NASA grants 
NNX09AD01G, NNX10AE83G, and NNX14AE62G. This research has made use of data and software provided by 
the High Energy Astrophysics Science Archive Research Center (HEASARC), 
which is a service of the Astrophysics Science Division at NASA/GSFC and the 
High Energy Astrophysics Division of the Smithsonian Astrophysical Observatory.

\bsp
\label{lastpage}

\end{document}